\documentclass[aps,prd,showpacs,eqsecnum,twocolumn,superscriptaddress,floatfix,nofootinbib]{revtex4-1}

\def\mnras{Mon.~Not.~R.~Astron.~Soc.}

\usepackage{amsmath}
\usepackage{latexsym}
\usepackage{times}
\usepackage{amssymb}
\usepackage{tabularx} 
\usepackage{hyperref}

\newcommand{\eg}{e.g.,~} 
\newcommand{\ie}{i.e.,~}
\newcommand{\cf}{cf.,~}

\usepackage{ifpdf}
\usepackage{graphicx}
\usepackage{subfigure}
\graphicspath{{./}}
\newcommand{\imgname}[1]{#1.eps}

\ifpdf
  \graphicspath{{./}}
  \renewcommand{\imgname}[1]{#1.pdf}
  
\fi

\begin{document} 

\title[Dynamical bar-mode instability in rotating and magnetized
  relativistic stars]{Dynamical bar-mode instability in rotating and
  magnetized relativistic stars}

\author{Luca \surname{Franci}}
\affiliation{Universit\`a di Parma and INFN gruppo collegato di Parma, Italy}
\author{Roberto \surname{De Pietri}}
\affiliation{Universit\`a di Parma and INFN gruppo collegato di Parma, Italy}
\author{Kyriaki \surname{Dionysopoulou}}
\affiliation{Max-Planck-Institut f\"ur Gravitationsphysik,
Albert-Einstein-Institut, Golm, Germany}
\author{Luciano \surname{Rezzolla}}
\affiliation{Max-Planck-Institut f\"ur Gravitationsphysik,
Albert-Einstein-Institut, Golm, Germany}
\affiliation{Institut f\"ur Theoretische Physik, Frankfurt am Main, Germany}

\date{\today}
\begin{abstract}
  We present three-dimensional simulations of the dynamical bar-mode
  instability in magnetized and differentially rotating stars in full
  general relativity. Our focus is on the effects that magnetic fields
  have on the dynamics and the onset of the instability. In particular,
  we perform ideal-magnetohydrodynamics simulations of neutron stars that
  are known to be either stable or unstable against the purely
  hydrodynamical instability, but to which a poloidal magnetic field in
  the range of $10^{14}$--$10^{16}$ G is superimposed initially. As
  expected, the differential rotation is responsible for the shearing of
  the poloidal field and the consequent linear growth in time of the
  toroidal magnetic field. The latter rapidly exceeds in strength the
  original poloidal one, leading to a magnetic-field amplification in the
  the stars. Weak initial magnetic fields, \ie $ \lesssim 10^{15}$ G,
  have negligible effects on the development of the dynamical bar-mode
  instability, simply braking the stellar configuration via
  magnetic-field shearing, and over a timescale for which we derived a
  simple algebraic expression. On the other hand, strong magnetic fields,
  \ie $\gtrsim 10^{16}$ G, can suppress the instability completely, with
  the precise threshold being dependent also on the amount of
  rotation. As a result, it is unlikely that very highly magnetized
  neutron stars can be considered as sources of gravitational waves via
  the dynamical bar-mode instability.
\end{abstract}


\maketitle

\section{Introduction}
\label{sec:intro}

Main-sequence stars with masses greater than about $8~M_{\odot}$ follow
two evolutionary paths; either they form a degenerate core of O/Ne/Mg, or
a degenerate Fe core, which, after undergoing a Type II supernova core
collapse, forms a proto-neutron star~\cite{Woosley02,Heger03}. Neutron
stars (NSs) are also expected to form through the accretion-induced
collapse of a white dwarf~\cite{Liu01,Fryer99}. At birth, NSs are rapidly
and differentially rotating, which makes them subject to various types of
instabilities. Among these, the dynamical bar-mode instability and the 
shear-instability are particularly interesting because of their potential
role as sources of gravitational waves (GWs).

Indeed, a newly born NS may develop a dynamical bar-mode instability when
the rotation parameter $\beta := T/ |W|$, with $T$ the rotational
kinetic energy and $W$ the gravitational binding energy, exceeds a
critical value $\beta_c$ (see, for instance,
~\cite{Stergioulas03,Andersson03} for some reviews). Under these
conditions, the rapidly rotating NS can become severely deformed, leading
to a strong emission of GWs in the kHz range. Analytic investigations of
the conditions under which these dynamical instabilities develop in
self-gravitating rotating stars can be found
in~\cite{Chandrasekhar69,Watts:2003nn}, but these are inevitably
restricted to Newtonian gravity or to simple shell models. To improve our
understanding of these instabilities also in the nonlinear regimes, and
to be able to extract useful physical information from the potential GW
emission, it is clear that a general-relativistic numerical modeling is
necessary. This has been the focus of a number of recent works, \eg
\cite{Shibata:2000jt,Baiotti06b,Manca07,Corvino:2010}, which have
provided important clues about the threshold for the instability and its
survival under realistic conditions. As an example, for a polytropic
relativistic star with polytropic index $\Gamma=2$, the calculations
reported in~\cite{Baiotti06b} revealed that the critical value is
$\beta_c\sim 0.254$, and that a simple dependence on the stellar
compactness can be used to track this threshold from the Newtonian limit
over to the fully relativistic one~\cite{Manca07}. Furthermore, numerical
simulations have also revealed that the instability is in general
short-lived and that the bar-deformation is suppressed over a timescale
of a few revolutions (this was first pointed out in
Ref.~\cite{Baiotti06b} and later confirmed in Ref.~\cite{Saijo2008},
where it was interpreted as due to a Faraday resonance).

One aspect of the bar-mode instability that so far has not received
sufficient attention is about the occurrence of the instability in
magnetized stars. This is not an academic question since NSs at birth are
expected to be quite generically magnetized, with magnetic fields that
have strengths up to $\simeq 10^{12}$ G in ordinary NSs and reaching
strengths in excess of $10^{15}$ G in magnetars, if instabilities or
dynamos have taken place in the proto-neutron star
phase~\cite{ThompsonDuncan1993,Bonanno:2003uw}.

Magnetic fields of this strength can affect both the structure and the
evolution of NSs~\cite{Lasky2011,Ciolfi2011,Ciolfi2012,Lasky2012}, and it
is natural to expect that they will influence also the development of the
instability when compared to the purely hydrodynamical case. A first
dynamical study in this direction has been carried out recently in
Ref. \cite{Camarda:2009mk}, where the development of the dynamical
bar-mode instability has been studied for differentially rotating
magnetized stars in Newtonian gravity and in the
ideal-magnetohydrodynamics (MHD) limit (\ie with a plasma having infinite
conductivity). Not surprisingly, this study found that magnetic fields do
have an effect on the development of the instability, but that this is
the case only for very strong magnetic fields. We here consider the same
problem, but extend the analysis to a fully general-relativistic
framework, assessing the impact that the results have on high-energy
astrophysics and GW astronomy.

Our investigation of the dynamics of highly-magnetized and rapidly
rotating NSs is also part of a wider study of this type of objects to
explain the phenomenology associated with short gamma-ray bursts. These
catastrophic phenomena, in fact, are normally thought to be related to
the merger of a binary system of NSs~\cite{Paczynski86, Eichler89,
  Narayan92, Rezzolla:2011}, which could then lead to the formation of a
long-lived hypermassive NS (HMNS)~\cite{Nakar:2007yr, Lee:2007js,
  Baiotti08, Rezzolla:2010}. If highly magnetized, the HMNS could then
also lead to an intense electromagnetic emission~\cite{Bucciantin2012a,
  Zhang2013}. This scenario has recently been considered in
Refs. \cite{Shibata2011b,Kiuchi2012b}, where numerical simulations of an
axisymmetric differentially rotating HMNS were carried out. The HMNS had
initially a purely poloidal magnetic field, which eventually led to a
magnetically driven outflow along the rotation axis.

A similar setup has also been considered in a number of works, either in
two-dimensional (2D) \cite{Duez:2006qe} or in three-dimensional (3D)
simulations \cite{Siegel2013}, with the goal of determining whether or
not the conditions typical of a HMNS can lead to the development of the
magnetorotational (MRI) instability
\cite{Velikhov1959,*Chandrasekhar1960}. Although this type of simulations
in 3D still stretches the computational resources presently available,
the very high resolutions employed in Ref.~\cite{Siegel2013}, and the
careful analysis of the results, provided the first convincing evidence
that the MRI can develop from 3D configurations. This has of course
important consequences on much of the phenomenology associated with
HMNSs, as it shows that very strong magnetic fields, up to equipartition,
will be produced in the HMNS if this survives long enough for the MRI to
develop.

In the simulations reported here we necessarily adopt much coarser
resolutions and hence we will not be able to concentrate our attention on
the development of the MRI. Rather, we will here extend our previous work
on the dynamical bar-mode instability~\cite{Baiotti06b, Manca07,
  DePietri06} also to the case of magnetized stars, determining when and
how magnetic fields can limit the development of the dynamical bar-mode
instability. Our initial models correspond to stationary equilibrium
configurations of axisymmetric and rapidly rotating relativistic
stars. More precisely, our initial models are described by a polytropic
EOS with adiabatic index $\Gamma = 2$ and are members of a sequence with
a constant rest-mass of $M \simeq 1.5 \; M_{\odot}$ and a constant amount
of differential rotation.

Interpreting the results of our simulations can be rather
straightforward. Because we work in the ideal-MHD limit, the magnetic
field lines are ``frozen'' in the fluid and follow its dynamics
(see~\cite{Dionysopoulou:2012pp} for a recent extension of the code to
resistive regime). As a consequence, differential rotation drives the
initial purely poloidal magnetic field into rotation, winding it up and
generating a toroidal component. At early times, the toroidal magnetic
field grows linearly with time, tapping the NS's rotational energy. At
later times, the growth starts deviating from the linear behavior and the
magnetic tension produced by the very large magnetic-field winding,
alters the angular velocity profile of the star. Depending on the models
adopted and the initial magnetic field strength, the magnetic winding
could become the most efficient mechanisms for redistributing angular
momentum, with the MRI being the dominant one when the Alfv\'en timescale
becomes comparable to the magnetic winding timescale.

Overall, we find that if the initial magnetic fields are $\lesssim
10^{15}$ G, then they have a negligible effect on the occurrence of the
dynamical bar-mode instability, which develops in close analogy with the
purely hydrodynamical case. On the other hand, if the initial magnetic
fields are $\gtrsim 10^{16}$ G, they can suppress the instability
completely. Note that the precise threshold marking the stability region
depends not only on the strength of the magnetic field, but also on the
amount of rotation. We trace this threshold by performing a number of
simulations of a number of sequences having the same parameter $\beta$
but different magnetizations. An important consequence of our results is
that because the instability is suppressed in strongly magnetized NSs,
these can no longer be considered as potential sources of GWs, at least
via the dynamical bar-mode instability. 

The organization of the paper is as follows. In Sect. \ref{sec:evolution}
we describe the numerical methods and the setup employed in our
simulations, as well as the full set of equations we solve. In
Sect. \ref{sec:initial} we mention briefly the main properties of the
stellar models adopted as initial data, together with the simplifications
and assumptions we make. In Sect. \ref{sec:results} we examine in great
detail the effects of the presence of an initial poloidal magnetic field
on differentially rotating stars covering a wide range in the parameter
space. We further discuss the qualitative and quantitative features of
the evolution of models with the same total rest mass but different
rest-mass density and angular momentum profiles that are known to be
unstable to the purely hydrodynamic bar-mode instability. Finally, we
investigate whether magnetic fields affect the stellar evolution even
when the bar-mode instability does not develop. Our conclusions are drawn
in Sect. \ref{sec:conclusions} and two appendices discuss the influence
of symmetries on the development of the instability and the convergence
of our results. Unless stated differently, we adopt geometrized units in
which $c = 1$, $G = 1$, $M_{\odot} = 1$.

\section{Mathematical and numerical setup}
\label{sec:evolution}

The simulations have been carried out using the general-relativistic
ideal-MHD (GRMHD) code \texttt{WhiskyMHD}~\cite{Giacomazzo:2007ti,
  Giacomazzo:2010, Rezzolla:2011}. The code provides a 3D numerical
solution of the full set of the GRMHD equations in flux-conservative form
on a dynamical background in Cartesian coordinates. It is based on the
same high-resolution shock-capturing (HRSC) techniques on domains with
adaptive mesh refinements
(AMR)~\cite{Schnetter:2003rb,carpet-author-and-web-site}) as discussed
in~\cite{Baiotti04}. The reconstruction method adopted is the one
discussed in the piecewise parabolic (PPM)~\cite{colella_1984_ppm}, while
the Harten-Lax-van Leer-Einfeldt (HLLE) approximate Riemann
solver~\cite{Harten83} has been employed in order to compute the
fluxes. The divergence of the magnetic field is enforced to stay within
machine precision by employing the flux-CD approach as implemented
in~\cite{Giacomazzo:2010}, but with the difference that we adopt as
evolution variable the vector potential instead of the magnetic
field. This method ensures the divergence-free character of the magnetic
field since the magnetic field is computed as the curl of the evolved
vector potential using the same finite-differencing operators as the ones
for computing the divergence of the magnetic field. 

Because of the gauge invariance of Maxwell equations, a choice needs to
be made and we have opted for the simplest one, namely, the algebraic Maxwell
gauge. This choice can introduce some spurious oscillations close to the
AMR boundaries in highly dynamical simulations, but this has not been the
case for the simulations reported here. On the other hand, it has allowed
us to keep the divergence of the magnetic field essentially nearly at machine
precision. A more advanced prescription has been also introduced recently
in Ref.~\cite{Etienne2012a}; this approach requires a certain amount of
tuning for optimal performance and will be considered in future
works. Additional information on the code can also be found in
Refs.~\cite{Giacomazzo:2007ti, Giacomazzo:2010}.

Furthermore, to remove spurious post-shock oscillations in the magnetic
field we add a fifth-order Kreiss-Oliger type of dissipation~\cite{Kreiss73} 
to the vector potential evolution equation with a dissipation parameter of
$0.1$. Finally, the evolution of the
gravitational fields is obtained through the \texttt{CCATIE} code, which
provides the solution of the conformal traceless formulation of the
Einstein equations~\cite{Pollney:2007ss}. The time integration of the
evolution equations is achieved through a third-order accurate
Runge-Kutta scheme. Essentially all of the simulations presented in this
paper use a 3D Cartesian grid with four refinement levels and with outer
boundaries located at a distance $\sim 150$ km from the center of the
grid. The finest resolution is $\Delta x \simeq 0.550$ km (between $40$
and $60$ points across the stellar radius, depending on the model) and
the coarsest extends up to about $\sim 150$ km, namely more than five
times the stellar radius. Unless stated differently, all of the
simulations discussed hereafter have been performed imposing a bitant
symmetry, \ie a reflection symmetry across the $z = 0$ plane.

For convenience we report here the full set of the evolution equations we
solve numerically which consists in the coupled systems of Einstein
and MHD equations, \ie
\begin{align}
R_{\mu\nu} -\frac{1}{2} g_{\mu\nu} R &= 8 \pi T_{\mu\nu} \,,  
\label{eq:evolution1}\\
\nabla_\mu T^{\mu\nu} &= 0 \,, 
\label{eq:evolution3a}\\
\nabla_\mu (\rho u^{\mu}) &= 0 \,, \
\label{eq:evolution3b}\\
\nabla_\mu ^\star\!F^{\mu\nu} &= 0 \,, 
\label{eq:evolution2a}\\
\nabla_\mu F^{\mu\nu} &= 4\pi \mathcal{J}^{\nu} \,,
\label{eq:evolution2b}
\end{align}
where $R_{\mu\nu}, g_{\mu\nu}$ and $R$ are the Ricci tensor, the metric
tensor and the Ricci scalar, respectively. On the electromagnetic side,
$F_{\mu\nu}$ is the Maxwell tensor, dual of the Faraday tensor $^*
F_{\mu\nu}$, $\mathcal{J}^{\mu}$ is the current four-vector, and on the 
matter side $\rho$ is
the rest-mass density, $u^{\mu}$ is the 4-velocity of the fluid satisfying the 
normalization condition $u_\mu u^\mu = -1$. The total 
energy-momentum tensor $T^{\mu\nu}$ is the 
linear combination of the contributions coming
from a perfect fluid, \ie $T^{\mu\nu}_{\mathrm{fl}}$, and from the
electromagnetic fields, \ie $T^{\mu\nu}_{\mathrm{em}}$
\begin{equation*}
T^{\mu\nu}=T^{\mu\nu}_{\mathrm{em}} + T^{\mu\nu}_{\mathrm{fl}}\, .
\end{equation*}
where 
\begin{align}
T^{\mu\nu}_{\text{fluid}} & := \rho h u^{\mu} u^{\nu} + p g^{\mu\nu}\,,\\
T^{\mu\nu}_{\text{em}} & := F^{\mu\sigma}{F^{\nu}}_{\sigma} 
   - \frac{1}{4} g^{\mu\nu} F_{\alpha\beta}F^{\alpha\beta}  \nonumber\\
  &= \left (u^{\mu} u^{\nu} + \frac{1}{2} g^{\mu\nu} \right) b^2 - b^{\mu} b^{\nu}\,,
\end{align}
In the expressions above we recall that $h = 1 +\epsilon + p/\rho$ is the
specific enthalpy, $\epsilon$ the specific internal energy. Hence, the energy 
density in the rest-frame of the
fluid is just $e = \rho(1+\epsilon)$. At the same
time, the four-vector $b^{\mu}$ represents the magnetic field as measured
in the comoving frame, so that the Maxwell and Faraday tensors are
expressed as (see \cite{Giacomazzo:2007ti, Giacomazzo:2010} for details)
\begin{align}
F^{\mu\nu} &= \epsilon^{\mu\nu\alpha\beta} u_{\alpha} b_{\beta}
=  n^{\mu} E^{\nu} - n^{\nu} E^{\mu} +
\epsilon^{\mu\nu\alpha\beta} B_{\alpha} n_{\beta}
\,, \\
^{*\!}F^{\mu\nu} &= b^{\mu}u^{\nu} - b^{\nu}u^{\mu} = 
n^{\mu} B^{\nu} - n^{\nu} B^{\mu} -
\epsilon^{\mu\nu\alpha\beta} E_{\alpha} n_{\beta}\,,
\end{align}
where the second equalities introduce the electric and magnetic fields
measured by an observer moving along a normal direction $n^{\nu}$. We
further note that the $\sqrt{4 pi}$ terms appearing in
Eqs.~{\eqref{eq:evolution2a},\eqref{eq:evolution2b}} are absorbed in
the definition of the magnetic field.

In the interest of compactness, we will not discuss here the detailed
formulation of the Eqs.~\eqref{eq:evolution1}--\eqref{eq:evolution3b} we
use in the numerical solution and refer the interested reader to the
following works where these aspects are discussed in detail:
Ref.~\cite{Pollney:2007ss} for the formulation of the Einstein equations
and the gauge conditions used, Refs.~\cite{Giacomazzo:2007ti,
  Giacomazzo:2010} for the formulation of the MHD equations and the
strategy for enforcing a zero divergence of the magnetic field,
Refs.~\cite{Baiotti08,Rezzolla_book:2013} for the computational
infrastructure and the numerical methods used. What is however important
to remark here is that we employ an ``ideal-fluid'' (or Gamma-law)
equation of state (EOS)~\cite{Rezzolla_book:2013}
\begin{equation}
\label{eq:EOSideal}
p = \rho\,\epsilon(\Gamma-1) \,,
\end{equation}
where $\Gamma$ is the adiabatic exponent, which we set to be $\Gamma=2$.
More realistic EOS could have been used, as done for instance in
Ref.~\cite{Galeazzi2013}, and this will indeed be the focus of future
work. At this stage, however, and because this is the first study of this
type, the simpler analytic EOS~\eqref{eq:EOSideal} is sufficient to
collect the first qualitative aspects of the development of the
instability.

\begin{table*}
\newcommand{\SKIP}{{~~~~~~~~~~~~~~}}
\newcommand{\comp}{{$M/R_e$}}
\newcommand{\rhoc}{{$\rho_{c}$}}
\newcommand{\Rrate}{{$r_{p}/r_{e}$}}
\newcommand{\msec}{{\small [ms]}}
\newcommand{\km}{{\small [km]}}
\newcommand{\rads}{{\small [rad/s]}}
\newcommand{\Msun}{{\small [$M_\odot$]}}
\newcommand{\expq}{~~{$\scriptstyle (10^{-4})$}~~}
\newcommand{\expD}{~~{$\scriptstyle (10^{-2})$}~~}
\newcommand{\expM}{~~{$\scriptstyle (10^{-6})$}~~}
\begin{tabular}{|c|ccc|cccc|cccc|cccc|}
\hline
\hline

Model &  \rhoc~~ &\Rrate & $A_b$ &$R_{e}$ & $M_{0}$ & $M$   & \comp  & $J$  & $J/M^2$  & $\Omega_{c}$ & $\Omega_{e}$ & $T$   & $W$   & $\beta$ & $\beta_{\mathrm{mag}}$ \\
      &  \expq   &\SKIP  &\SKIP  & \km    & \Msun   & \Msun & \SKIP  &\SKIP &\SKIP & \rads        & \rads        & \expD & \expD & \SKIP   & \expM \\
\hline 
\texttt{U13} & $0.599$ & $0.200$ & $1.85\times 10^{6}$ &  $35.9$ & $1.505$ & $1.462$ & $0.0601$ & 3.747 & $1.753$ & $3647$ & $1607$ & $2.183$ &$ 7.764$ & $0.2812$ & $5.3$ \\
\texttt{U11} & $1.092$ & $0.250$ & $1.46\times 10^{6}$ &  $34.4$ & $1.507$ & $1.460$ & $0.0627$ & 3.541 & $1.661$ & $3997$ & $1747$ & $2.284$ &$ 8.327$ & $0.2743$ & $4.7$ \\
\texttt{U3 } & $1.672$ & $0.294$ & $8.74\times 10^{5}$ &  $32.4$ & $1.506$ & $1.456$ & $0.0664$ & 3.261 & $1.538$ & $4434$ & $1916$ & $2.352$ &$ 9.061$ & $0.2596$ & $3.5$ \\
\hline                                                                                                                                                
\texttt{S1 } & $1.860$ & $0.307$ & $6.94\times 10^{5}$ &  $31.6$ & $1.512$ & $1.460$ & $0.0682$ & 3.191 & $1.497$ & $4593$ & $1976$ & $2.384$ &$ 9.388$ & $0.2540$ & $3.0$ \\
\texttt{S6 } & $2.261$ & $0.336$ & $4.50\times 10^{5}$ &  $30.0$ & $1.505$ & $1.449$ & $0.0713$ & 2.965 & $1.412$ & $4901$ & $2093$ & $2.369$ &$ 9.859$ & $0.2403$ & $2.3$ \\
\texttt{S7 } & $2.754$ & $0.370$ & $2.01\times 10^{5}$ &  $28.1$ & $1.506$ & $1.447$ & $0.0760$ & 2.741 & $1.309$ & $5284$ & $2234$ & $2.360$ &$ 10.56$ & $0.2234$ & $1.0$ \\
\texttt{S8 } & $3.815$ & $0.443$ & $5.96\times 10^{4}$ &  $26.7$ & $1.506$ & $1.439$ & $0.0862$ & 2.322 & $1.121$ & $5995$ & $2482$ & $2.255$ &$ 11.96$ & $0.1886$ & $0.4$ \\
\hline
\hline
\end{tabular}
\caption{Main properties of the stellar models evolved in the
  simulations. In the first column we report the model name, while in the
  next three the parameters we used to generate the initial models,
  namely the central rest-mass density $\rho_c$, the ratio between the
  polar and the equatorial coordinate radii $r_{p}/r_{e}$ and the
  parameter $A_b$ of Eq.~(\ref{eq:defA_b}) that would generate a magnetic
  field whose initial maximum value in the ($x,y$) plane is $1 \times
  10^{15}$ G. In the remaining columns we report, from left to right, the
  proper equatorial radius $R_{e}$, the rest mass $M_{0}$, the
  gravitational mass $M$, the compactness $M/R_e$, the total angular
  momentum $J$, $J/M^2$, the angular velocities at the axis
  $\Omega_{c}=\Omega(r=0)$ and at the equator $\Omega_{e}=\Omega(r=R_e)$,
  the rotational kinetic energy $T$ and the gravitational binding energy
  $W$, their ratio $\beta=T/|W|$ (instability parameter) and finally the
  ratio between the total magnetic energy and the sum of the rotational
  energy and the gravitational binding energy
  ($\beta_{\rm{mag}}=E_{\text{mag}}/(T+|W|)$). Unless explicitly stated,
  all these quantities are expressed in geometrized units in which
  $G=c=M_\odot=1$.}
\label{table:models}
\end{table*}

\section{Initial data and diagnostics}
\label{sec:initial}
\begin{figure*}
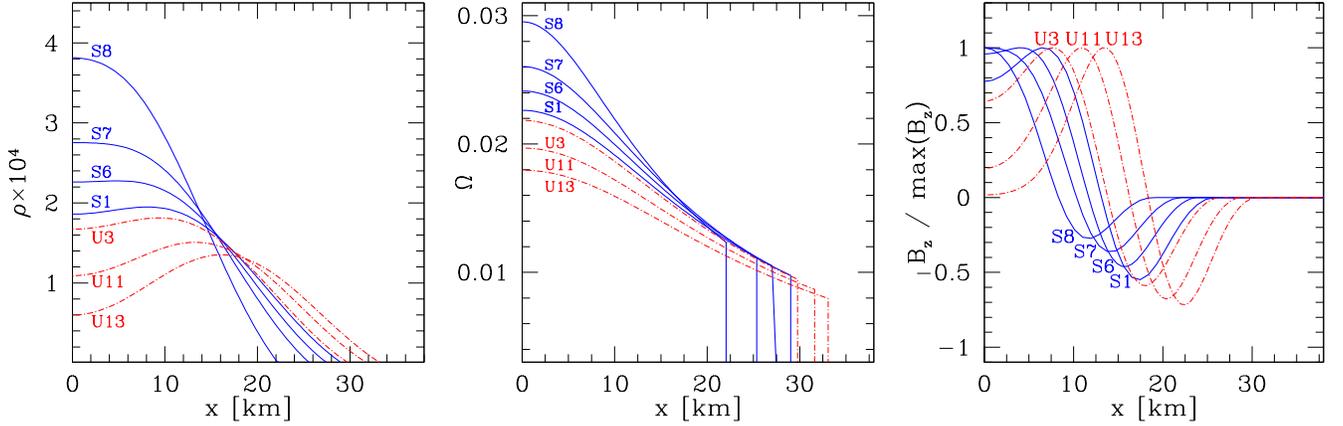

\begin{center}
\includegraphics[width=5.9cm]{\imgname{rho_vs_x_0}}
\includegraphics[width=5.9cm]{\imgname{omega_vs_x_0}}
\includegraphics[width=5.9cm]{\imgname{Bz_vs_x_0}} 
\end{center}
\caption{Initial profiles of the rest-mass density $\rho$ (left panel),
  the angular velocity $\Omega$ (center panel) and of the $z$-component
  of the magnetic field (right panel) for models \texttt{S8},
  \texttt{S7}, \texttt{S6}, \texttt{S1}, \texttt{U3}, \texttt{U11} and
  \texttt{U13}. The profiles of the stable models are here drawn with
  blue solid lines, while those for the unstable models with red
  dot-dashed lines.}
\label{fig:InitialProfile}
\end{figure*}

The initial data of our simulations are computed as stationary
equilibrium solutions of axisymmetric and rapidly rotating relativistic
stars in polar coordinates and without magnetic
fields~\cite{Stergioulas95}. In generating these equilibrium models we
adopt a ``polytropic'' EOS~\cite{Rezzolla_book:2013}, $p = K
\rho^{\Gamma}$, with $K=100$ and $\Gamma=2$, and assume the line element
for an axisymmetric and stationary relativistic spacetime to have the
form
\begin{equation} 
\begin{split}
ds^{2} = - e^{\mu+\nu} dt^{2}
         + e^{\mu-\nu} r^{2} \sin^{2}\theta (d\phi-\omega dt)^{2}
\\
         +e^{2\xi}(dr^{2}+r^{2} d\theta^{2})\,,
\end{split}
\end{equation} 
where $\mu$, $\nu$, $\omega$ and $\xi$ are space-dependent metric
functions. To reach the large angular momentum needed to trigger the
dynamical bar-mode instability, a considerable amount of differential
rotation needs to be introduced and we do so following the traditional
constant specific angular momentum law (``j-constant'') of differential 
rotation, in which the angular
velocity distribution takes the form~\cite{Komatsu89,Friedman2012}
\begin{equation} 
\Omega_{c} -\Omega   =   \frac{1}{\hat{A}^{2} R_{e}^{2} }
       \left[ \frac{(\Omega-\omega) r^2   \sin^2\theta e^{-2\nu}
              }{1-(\Omega-\omega)^{2} r^2 \sin^2\theta e^{-2\nu}
       }\right] \,,
\label{eq:velocityProfile}
\end{equation} 
where $R_{e}$ is the coordinate equatorial stellar radius and the
coefficient $\hat{A}$ is a measure of the degree of differential
rotation, which we set to $\hat{A}=1$ in analogy with other works in
the literature. Once imported onto the Cartesian grid and throughout
the evolution, we compute the angular velocity $\Omega$ (and the
period $P$) on the $(x,y)$ plane as
\begin{equation}
\Omega :=\frac{u^{\phi}}{u^{0}}
       =\frac{u^{y} \cos \phi - u^{x} \sin \phi}{u^{0}\sqrt{x^2+y^2}}
       \, ,
       \qquad P=\frac{2\pi}{\Omega}\,.
\end{equation}
Other characteristic quantities of the system, such as the baryon mass
$M_0$, the gravitational mass $M$, the angular momentum $J$, the
rotational kinetic energy ${T}$, and the gravitational binding energy
${W}$ are calculated as in~\cite{Stergioulas98}
\begin{eqnarray}
\label{relevantquantities}
M   & := & \int\!d^3\!x\, \alpha \sqrt{\gamma} \left[
  -2(T_{\mathrm{fl}})^0_{\ 0} + 
(T_{\mathrm{fl}})^\mu_{\ \mu} \right] 
\,,  \label{eq:DEF M}\\
M_0 & := & \int\!d^3\!x\,  \sqrt{\gamma} D 
\,, \label{eq:DEF M0}\\
{E_{\rm int}} & := & \int\!d^3\!x\, \sqrt{\gamma} D \epsilon
\,, \label{eq:DEF Ein}\\
J   & := & \int\!d^3\!x\, \alpha \sqrt{\gamma} (T_{\mathrm{fl}})^0_{\ \phi} 
\,, \label{eq:DEF J}\\
{T} & := & \frac{1}{2} \int\!d^3\!x\, \alpha \sqrt{\gamma} \Omega  
(T_{\mathrm{fl}})^0_{\ \phi} 
\,, \label{eq:DEF T}\\
{W} & := & {T} + {E_{\rm int}} + M_0 - M 
\,, \label{eq:DEF W} 
\end{eqnarray}
where $\epsilon$ is the specific internal energy, $D$ is the conserved
rest-mass density, $\gamma$ is the determinant of the three-metric and
$(T_{\mathrm{fl}})^{\mu}_{\ \nu}$ corresponds to the fluid contributions
to the stress-energy tensor. A couple of important caveats need to be
made about the definitions above. First, we note that we have defined the
gravitational mass and angular momentum taking into account only the
fluid part of the energy-momentum tensor and thus neglecting the
electromagnetic contributions. This is strictly speaking incorrect, but
tolerable given that the relative electromagnetic contributions to the
mass and angular momentum are $\lesssim 10^{-5}$. Second, the definitions
above for $J$, ${T}$, ${W}$ and $\beta$ are meaningful only in the case
of stationary axisymmetric configurations and should therefore be treated
with care once the rotational symmetry is lost.

\begin{figure}
\noindent
\includegraphics[width=0.48\textwidth]{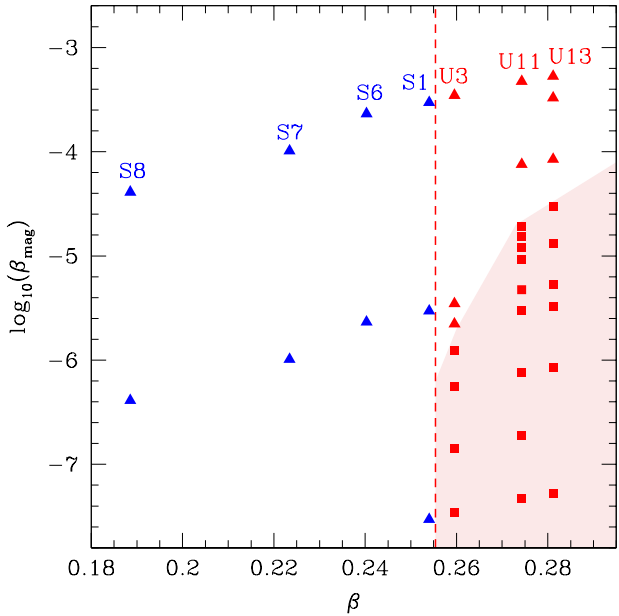}
\caption{Representation of the initial models in a
  $(\beta,\beta_{\rm{mag}})$ plane. Blue and red symbols mark models
  that are respectively bar-mode stable and bar-mode unstable at zero
  magnetizations, while the vertical red dashed line marks the
  stability threshold for zero magnetic fields. The red-shaded area
  collects as a function of their magnetization models that the
  evolutions reveal to be bar-mode unstable; hence, red squares refer
  to initial models that develop a bar-mode instability, while red
  triangles refer to potentially bar-unstable models that are
  stabilized by the strong magnetic fields.}
\label{fig:InitialModels}
\end{figure}

The main properties of all the stellar models we have used as initial
data are reported in Table~\ref{table:models}, where we have introduced
part of our notation to distinguish the different models. In particular
models indicated as \texttt{U*} and as \texttt{S*} refer to NSs that are
unstable and stable to the purely dynamical bar-mode instability,
respectively (this result was determined in
Refs.~\cite{Baiotti06b,Manca07}).  Figure~\ref{fig:InitialProfile} shows
the initial profiles of the rest-mass density $\rho$ (left panel), of the
rotational angular velocity $\Omega$ (central panel), and of the
$z$-component of the magnetic field (right panel) for all the models we
have evolved. The profiles for the models that are unstable in the
unmagnetized case are drawn with blue solid lines, while we use red
dot-dashed lines for stable models. Note that the position of the maximum
of the rest-mass density coincides with the center of the star only for
models with low $\beta$; for those with a larger $\beta$, the maximum of
the rest-mass density resides, instead, on a circle on the equatorial
plane.

All the equilibrium models are members of a sequence having a constant
rest-mass $M_0 \simeq 1.5\ M_{\odot}$ and are stable to gravitational
collapse on the basis of the results of~\cite{Giacomazzo2011}. An initial
poloidal magnetic field is added as a perturbation to the initial
equilibrium models by introducing a purely toroidal vector potential
$A_\phi$ given by
\begin{equation}
A_\phi = A_b \left( \max(p - p_{\mathrm{cut}},0) \right)^2\,,
\label{eq:defA_b}
\end{equation} 
where $p_{\mathrm{cut}}$ is 4 \% of the maximum pressure, while $A_b$ is
chosen in a way to have the chosen value for the maximum of the initial
magnetic field $B$. The Hamiltonian and momentum constraint equations are
not solved after superimposing the magnetic field, but we have
verified that for the magnetic-field strength considered here, this
perturbation introduces only negligible additional violations of the
constraints.

The strength of the initial magnetic field can be characterized by the
value of the ratio between the total magnetic energy
\begin{equation}
{E_{\rm mag}} := \int\!d^3\!x\,\alpha^2\sqrt{\gamma}\, 
T^{00}_{\textrm{em}}\,,
 \label{eq:DEF Em}
\end{equation}
and the sum of the rotational kinetic energy $T$ and of the gravitational
binding energy $W$, which we indicate as $\beta_{\mathrm{mag}}:=
E_{\rm{mag}}/(T+|W|)$, in analogy with the instability parameter
$\beta:= T/|W|$. This parameter should not be confused with what is
usually defined as the $\beta$ parameter of a plasma, \ie the ratio of the 
fluid pressure to the magnetic pressure.

In Table~\ref{table:models} we also report the values of the coefficient
$A_b$ [see Eq.~\eqref{eq:defA_b}] and the parameter
$\beta_{\mathrm{mag}}$ corresponding to an initial poloidal magnetic field
strength equal to $10^{15}$ G. All the initial models are also
reported in Fig. \ref{fig:InitialModels} according to the values of their
parameters $\beta$ and $\beta_{\mathrm{mag}}$. The models that are known
to be stable against the bar-mode instability in the unmagnetized case
are here drawn in blue (\texttt{S1}, \texttt{S6}, \texttt{S7} and
\texttt{S8}), while the unstable ones are drawn in red (\texttt{U3},
\texttt{U11} and \texttt{U13}). The different symbols used in this figure
will be further discussed in Sect. \ref{sec:results} when illustrating
the results of our work; it is sufficient to say for now that squares and
triangles refer to unstable models with unmodified and modified growth
times, respectively. Hereafter we will also extend our notation and
denote a particular magnetized model by marking it by the maximum
initial value of the $z$-component of the magnetic field on the $(x,y)$
plane, \ie $B^{z}_\text{max}|_{t,z=0}$, expressed in Gauss. As an
example, the bar-mode unstable model with initial
$B^{z}_\text{max}|_{t,z=0}=1.0 \times 10^{15}$ G will be indicated as
\texttt{U11-1.0e15}.

In order to analyze better the effects of magnetic fields on the dynamics
of the bar-mode instability, we have introduced additional diagnostic
variables to quantify and describe the evolution of the magnetic
field itself. For axisymmetric configurations one usually decomposes the
magnetic field in toroidal and poloidal components, studying their
dynamics separately. When axisymmetry is lost, however, this nice
decomposition is no longer available. Nevertheless, there exists a
decomposition that can be defined even if axisymmetry is not preserved,
which is reduced to the usual poloidal-toroidal one in the axisymmetric
stationary case. The main idea of this decomposition is to separate the
magnetic field in a component in the direction of the fluid motion and
hence parallel to the fluid three-velocity and in a component that is
orthogonal to it.  We therefore split the magnetic field measured by an
Eulerian observer as
\begin{equation}
  B^i = B_{\parallel} \frac{v^i}{\sqrt{\gamma_{ij}v^i v^j}} + B_{\perp}^i\,,
\end{equation}
where we define the ``perpendicular'' part of the magnetic field from the
condition $B_{\perp}^i v_i =0$, while the ``parallel'' part is a scalar
and is trivially defined as $B_{\parallel} :=
B^jv_j/(v^iv_i)^{1/2}$. Initially, when the flow is essentially
azimuthal, $B^i_{\perp}$ corresponds to the poloidal component of the
magnetic field, while $B_{\parallel}v^i/(v^j v_j)^{1/2}$ to the toroidal
component. Hereafter we will refer loosely to these as the ``poloidal''
and ``toroidal'' components, respectively.

Within this decomposition, we can then define the electromagnetic
energy contributions associated to the ``toroidal'' and ``poloidal''
magnetic-field components as
\begin{eqnarray}
E_{\rm{mag}}^{\rm{tor}} &:=& \int\!d^3\!x\, \sqrt{\gamma}\,
\frac{1}{2} B_{\parallel} B_{\parallel}\,,
\label{eq:EMAG_tor}\\ 
E_{\rm{mag}}^{\rm{pol}} &:=& \int\!d^3\!x\, \sqrt{\gamma}\,
\frac{1}{2} \gamma_{ij} B_{\perp}^i B_{\perp}^j (1+\gamma_{rs}v^r v^s)\,.
\label{eq:EMAG_pol}
\end{eqnarray}
Note that the total electromagnetic energy satisfies the condition
$E_{\rm{mag}}=E_{\rm{mag}}^{\rm{tor}}+E_{\rm{mag}}^{\rm{pol}}$, since the
electric field $E^i$ provides a contribution to the energy, $E^iE_i =
(v^iv_i)(B^iB_i-B_{\parallel}^2)$, that is already included in the
definitions~\eqref{eq:EMAG_tor} and \eqref{eq:EMAG_pol}. Another
important set of diagnostic quantities focuses instead on the detection
of a bar deformation, which can be conveniently quantified in terms of
the distortion parameters~\cite{Saijo:2000qt}
\begin{eqnarray}
\eta_{+}      &:=\dfrac{{I}^{xx}-{I}^{yy}}{{I}^{xx}+{I}^{yy}} 
\,, \label{etap}\\
\eta_{\times} &:=\dfrac{2\; {I}^{xy}}{{I}^{xx}+{I}^{yy}}  
\,, \label{etac} \\
\eta          &:=\sqrt{\eta_{+}^{2}+\eta_{\times}^{2}} 
\,, \label{eq:Qdistortion} 
\end{eqnarray}
where the quadrupole moment of the matter distribution can be computed in
terms of the conserved density $D$ as in \cite{Baiotti06b,Baiotti07}
\begin{equation}
\label{eq:defQuadrupole}
I^{jk} = \int\! d^{3}\!x \; \sqrt{\gamma} D \; x^{j} x^{k} \,.
\end{equation}
Note that all quantities in Eqs.~(\ref{etap})--(\ref{eq:Qdistortion}) are
expressed in terms of the coordinate time $t$ and do not represent
therefore invariant measurements at spatial infinity. However, for the
simulations reported here, the length-scale of variation of the lapse
function at any given time is always larger than twice the stellar
radius at that time, ensuring that the events on the same time-slice
are also close in proper time.

In addition, $\eta_{+}$ can be conveniently used to quantify both the
growth time $\tau_{_{\rm bar}}$ of the instability and the oscillation
frequency $f_{_{\rm bar}}$ of the unstable bar once the instability is
fully developed. In practice, we obtain a measurement of $\tau_{_{\rm
    bar}}$ and $f_{_{\rm bar}}$ by performing a nonlinear least-square
fit of the computed distortion $\eta_{+}(t)$ with the trial function
\begin{equation}  
\eta_{+}(t) = \eta_{0} \; e^{t/\tau_{_{\rm B}}} 
                     \cos(2\pi\, f_{_{\rm B}} \, t+\phi_{0}) 
\,.
\label{eq:etafit}
\end{equation}  
%

\section{Results}
\label{sec:results}

\subsection{Effects of the magnetic field on unstable models}
\label{subsec:UntableModels}
We start by discussing in detail the results relative to model
\texttt{U11} when evolved for different values of the initial poloidal
magnetic field. The dynamics of this unstable model are very clear and
allow us to show a full qualitative and quantitative picture of what
happens as the bar-mode instability develops. We will therefore focus
our attention on models \texttt{U11-1.0e14}, \texttt{U11-2.0e15},
\texttt{U11-4.0e15} and \texttt{U11-1.0e16}, which, as discussed
before, have initial poloidal magnetic field such that
$B^{z}_\text{max}|_{t,z=0}$ is equal to $ 1.0 \times 10^{14}$, $2.0
\times 10^{15}$, $4 \times 10^{15}$ and $1.0 \times 10^{16}$ G,
respectively.

\begin{figure}
\begin{center}
\includegraphics[width=0.48\textwidth]{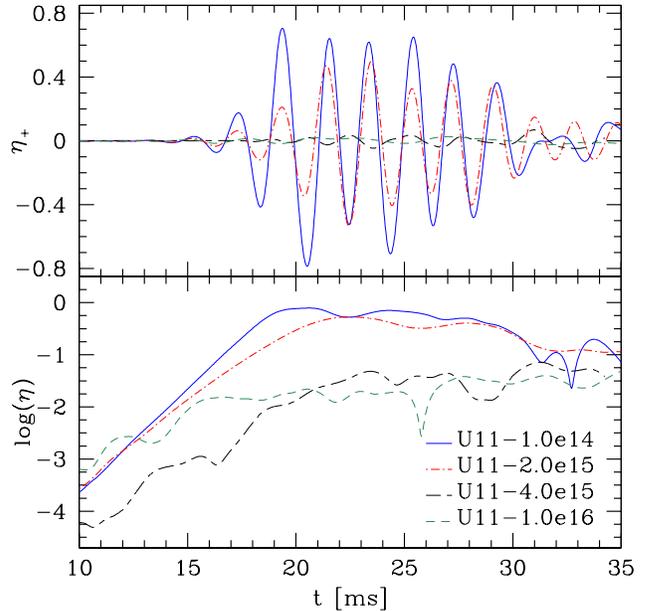}
\end{center}
\caption{Evolution of the distortion parameters $\eta_+$ and $\eta$ for
  model \texttt{U11} with four different values of the initial poloidal
  magnetic field: $B^{z}_\text{max}|_{t,z=0} = 1.0 \times 10^{14}$, $2.0
  \times 10^{15}$, $4.0 \times 10^{15}$, and $1.0 \times 10^{16}$ G.}
\label{fig:distorsion_bfield}
\end{figure}

\begin{figure*}
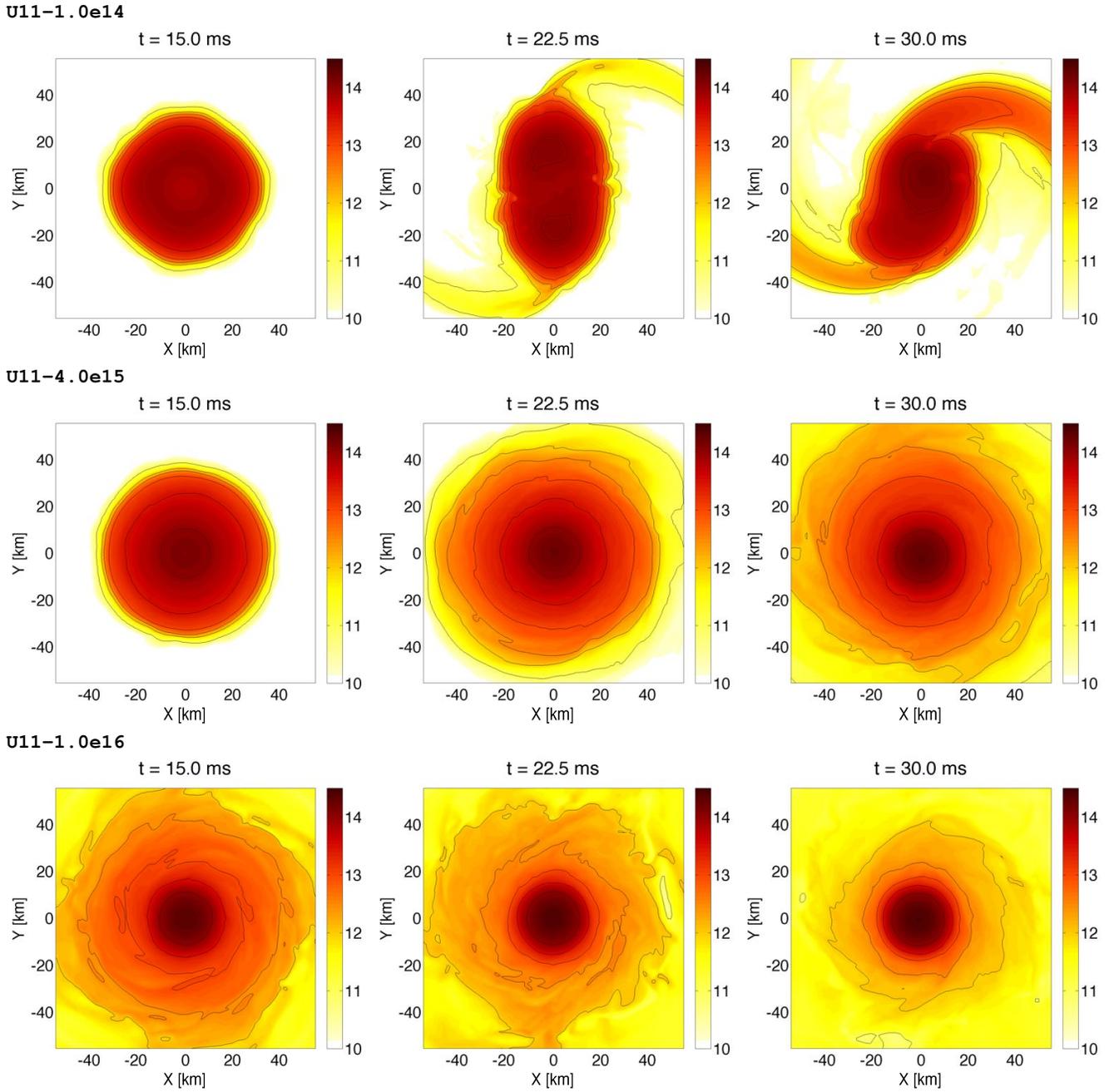

\begin{tabular}{ccc}
\multicolumn{3}{l}{\bf \texttt{U11-1.0e14}}\\
          \includegraphics[width=0.32\textwidth]{\imgname{U11B1e14_rho_151}}
&         \includegraphics[width=0.32\textwidth]{\imgname{U11B1e14_rho_225}}
&         \includegraphics[width=0.32\textwidth]{\imgname{U11B1e14_rho_300}}
\\
\multicolumn{3}{l}{\bf \texttt{U11-4.0e15}}\\
          \includegraphics[width=0.32\textwidth]{\imgname{U11B4e15_rho_151}}
&         \includegraphics[width=0.32\textwidth]{\imgname{U11B4e15_rho_225}}
&         \includegraphics[width=0.32\textwidth]{\imgname{U11B4e15_rho_300}}
\\
\multicolumn{3}{l}{\bf \texttt{U11-1.0e16}}\\
          \includegraphics[width=0.32\textwidth]{\imgname{U11B1e16_rho_151}}
&         \includegraphics[width=0.32\textwidth]{\imgname{U11B1e16_rho_225}}
&         \includegraphics[width=0.32\textwidth]{\imgname{U11B1e16_rho_300}}
\\
\end{tabular}
    \caption{Snapshots of the rest-mass density on the $(x,y)$ plane for
      model \texttt{U11-1.0e14} (top row), \texttt{U11-4.0e15} (central
      row) and \texttt{U11-1.0e16} (bottom row) at different times during
      the evolution, namely, $t = 15.0$ ms (left column), $t=22.5$ ms
      (central column) and $t=30$ ms (right column). Additionally,
      isodensity contours are shown for $\rho = 10^{6}, \; 10^{11}, \;
      10^{12}, \; 10^{12.5}, \; 10^{13}, \;10^{13.5}$, and $10^{14} \;
      \text{g~cm}^{-3}$.}%
   \label{fig:RhoSnaphots}
\end{figure*}

In Fig.~\ref{fig:distorsion_bfield} we show the evolution of the
distortion parameters $\eta_+$ (top panel) and $\eta$ (bottom panel) for
these models. In the least magnetized model (\ie \texttt{U11-1.0e14}),
$\eta_+$ starts oscillating after about 10 ms of evolution with an
amplitude that almost reaches unity, and it keeps oscillating for about
20 ms. At the same time, $\eta$ undergoes an exponential growth,
increasing its value by about three orders of magnitude until it reaches
a saturation level, which persists for about 10 ms and then decays. This
is exactly the behavior we expect from a stellar model which is unstable
against the dynamical bar-mode instability, as model \texttt{U11} is
known to be in the un-magnetized case (\cf
Refs.~\cite{Baiotti06b,Baiotti07}).

However, when the initial poloidal magnetic field is two orders of
magnitude stronger (\ie as for model \texttt{U11-1.0e16}), the dynamics
shows a very different behavior. The amplitude of the oscillations in
$\eta_+$ is negligible and $\eta$ does not grow exponentially, being two
orders of magnitude lower than it is for model \texttt{U11-1.0e14} during
the whole evolution. This indicates that although the model is unstable
in the absence of magnetic fields, no bar-mode deformation develops in
this case over a timescale of $\sim 35$ ms of evolution and for this
magnetic-field strength.

For intermediate initial poloidal magnetic fields, we find a
significant change in the dynamics by simply varying the field
strength by a factor of two, which corresponds to a change of a factor
of four in the magnetic energy. Moreover, in model \texttt{U11-2.0e15}
the bar-mode instability still develops, even though it takes a little
longer to grow, while model \texttt{U11-4.0e15} is stable and the
bar-mode instability is suppressed, since $\eta$ does not show an
exponential growth. As a result, we can bracket the stability
threshold for the development of the bar-mode instability between
these two models in the presence of strong magnetic fields (\cf
Fig.~\ref{fig:InitialModels}).

To better illustrate the different behavior of the matter evolution
for different initial poloidal magnetic field strengths, in
Fig.~\ref{fig:RhoSnaphots} we show three snapshots of the evolution of
the rest-mass density on the $(x,y)$ plane for three of the above
models (\ie models \texttt{U11-1.0e14}, \texttt{U11-4.0e15} and
\texttt{U11-1.0e16}) at times $t=15.0,\ 22.5,\ 30.0$ ms. In
particular, in the top row of Fig.~\ref{fig:RhoSnaphots} we show the
evolution of model \texttt{U11-1.0e14}, which as discussed previously
is bar-mode unstable as also its un-magnetized counterpart. After 15
ms we can already observe a small deformation with respect to the
initial axisymmetric configuration, which is then amplified until a
bar is fully formed after about 10 ms of the first oscillations
observed in $\eta_+$. In the central row we show the evolution of
model \texttt{U11-4.0e15}, which as mentioned before is instead stable
against bar-mode deformations due to the presence of the strong
magnetic field. In this case, after 15 ms the density profile has
already changed, turning from an initial toroidal profile (\cf
Fig.~\ref{fig:InitialProfile}) to an oblate profile with its maximum
residing on the $z$-axis. Later in the evolution, we observe an
increase in the central density and the outer layers expanding well
beyond the borders of the finest grid. Finally, on the bottom row we
show snapshots of the density for model \texttt{U11-1.0e16}, which
refers to the strongly magnetized case and which is also stable and
shows a similar behavior to the previous model. The only important
difference is the larger increase of the central rest-mass density and
the more significant expansion of the outer layers of the
star. Indeed, after the first 15 ms of evolution, matter has been shed
already beyond the edges of the finest grid.

\begin{figure*}
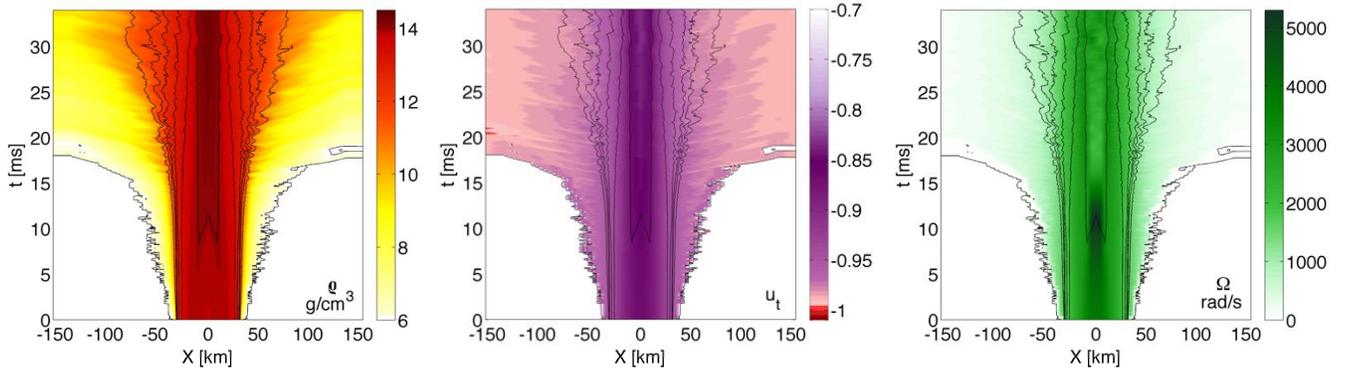
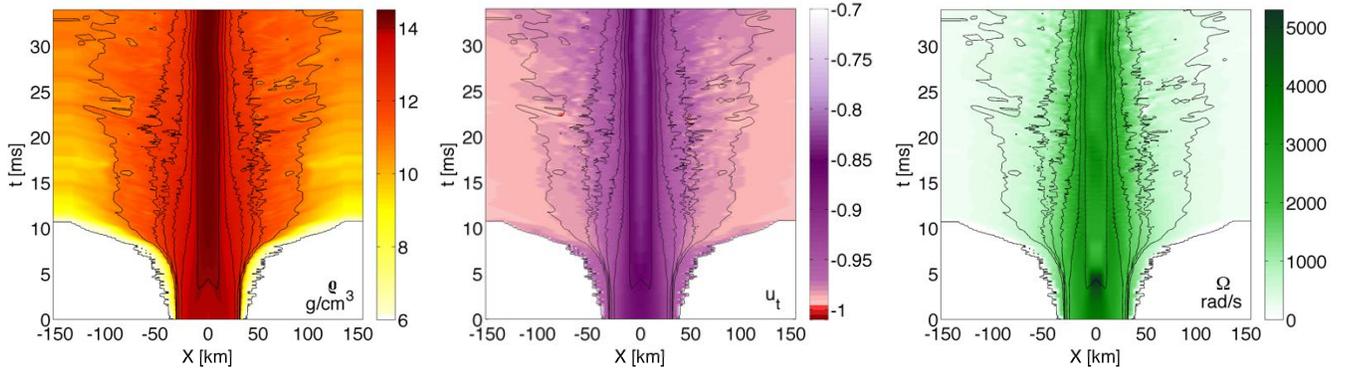

\begin{tabular}{ccc}
\multicolumn{3}{l}{\bf \texttt{U11-1.0e14}}\\[-1mm]
      \includegraphics[height=4.9cm]{\imgname{rho_U11B1e14}}
&     \includegraphics[height=5.2cm]{\imgname{ut_U11B1e14}}
\hspace{-0.5cm}
&     \includegraphics[height=4.9cm]{\imgname{Omega_U11B1e14}}
\\
\multicolumn{3}{l}{\bf \texttt{U11-4.0e15}}\\[-1mm]
      \includegraphics[height=4.9cm]{\imgname{rho_U11B4e15}}
&     \includegraphics[height=5.2cm]{\imgname{ut_U11B4e15}}
\hspace{-0.5cm}
&     \includegraphics[height=4.9cm]{\imgname{Omega_U11B4e15}}
\\
\multicolumn{3}{l}{\bf \texttt{U11-1.0e16}}\\[-1mm]
      \includegraphics[height=4.9cm]{\imgname{rho_U11B1e16}}
&     \includegraphics[height=5.2cm]{\imgname{ut_U11B1e16}}
\hspace{-0.5cm}
&     \includegraphics[height=4.9cm]{\imgname{Omega_U11B1e16}}
\\
\end{tabular}
\caption{Spacetime diagrams of the evolution of the rest-mass density
  $\rho$ (left column), of the time-component of the fluid
  four-velocity $u_t$ (central column), and of the angular velocity
  $\Omega$ (right column) along the $x$-axis. The models considered
  here are \texttt{U11-1.0e14} (top row), \texttt{U11-4.0e15} (central
  row), and \texttt{U11-1.0e16} (bottom row). The color code is
  indicated to the right of each plot. In addition, all diagrams also
  report isodensity contours of the rest-mass density $\rho = 10^{6},
  \; 10^{11}, \; 10^{12}, \; 10^{12.5}, \; 10^{13}, \;10^{13.5}$, and
  $10^{14} \; \text{g~cm}^{-3}$.}
\label{fig:spacetime_U11}
\end{figure*}

A deeper insight in the matter dynamics in the three different cases
discussed above can be gained through the spacetime diagrams shown
Fig.~\ref{fig:spacetime_U11}, and that are reminiscent of similar ones
first presented in Ref.~\cite{Rezzolla:2010}. In particular, the left
column of Fig.~\ref{fig:spacetime_U11} shows the rest-mass density
profile along the $x$-axis for the three models \texttt{U11} using both a
colormap (see the right-edge of the different panels) and some
representative contour lines; note that the colorcode and the color ranges
are the same in the three cases. It is worth mentioning that the
low-magnetic-field model \texttt{U11-1.0e14} (top panel in left column)
shows the evolution we expect from a bar-mode unstable model, since the
bar deformation is clearly visible after about 20 ms. The highly
magnetized model \texttt{U11-4.0e15} (middle panel in left column), on
the other hand, shows no bar deformation and exhibits instead a
transition from a toroidal configuration to an oblate one as is evident
in Fig.~\ref{fig:RhoSnaphots}. In addition, a small amount of matter is
shed on the equatorial plane after about 15 ms of evolution. Finally, for
the very highly magnetized model \texttt{U11-1.0e16} (bottom panel in
left column), the expansion of the outer layers is much more rapid and
the stellar material reaches a size of about 100 km (not shown in the
figure), which is almost twice as large as for model
\texttt{U11-4.0e15}. The ejected material creates an extended and
flattened envelope of high-density matter\footnote{It is tempting and
  sometimes encountered in the literature to refer to the envelope as
  ``disk'' or ``torus''; however, we find this is very misleading as the
  envelope is not disjoint from the star but rather an integral part of
  it which should not be discussed separately.}, with
rest-mass densities as high as $10^{12}~\textrm{g~cm}^{-3}$.

To determine whether the ejected matter is gravitationally bound or
not, we look at the time component of the fluid four-velocity $u_t$
(central column of Fig.~\ref{fig:spacetime_U11}) since the local
condition $u_t > -1$ provides a necessary although not sufficient
condition for a fluid element to be unbound \cite{Rezzolla:2010}. We
recall that this condition is exact only in an axisymmetric and
stationary spacetime. These requirements are not matched during the
matter-unstable phase, but the conditions can be used nevertheless as
a first approximation to determine whether part of the material is
actually escapes to infinity during the evolution. As is evident from
Fig.~\ref{fig:spacetime_U11}, this condition is fulfilled throughout
the whole evolution for the highly magnetized models
\texttt{U11-1.0e16} and \texttt{U11-4.0e15} not only on the finest
refinement level shown in Fig.~\ref{fig:spacetime_U11}, but on the
whole computational domain. However, this is not the case for model
\texttt{U11-1.0e14} at the time the bar-mode instability is fully
developed. In fact, in this case we observe that a certain amount of
unbound matter is shed in correspondence with the spiral arms of the
bar. The ejection of matter occurs only in very low-density regions
around the star, where $\rho \simeq 10^{10}~\textrm{g~cm}^{-3} \simeq
10^{-4} \, \rho_c$. Overall, the total amount of matter (both bound
and unbound) escaping from the outer grid after $20$ ms of evolution
is less than $0.2 \%$ of the total initial rest mass of the NSs.

\begin{figure*}
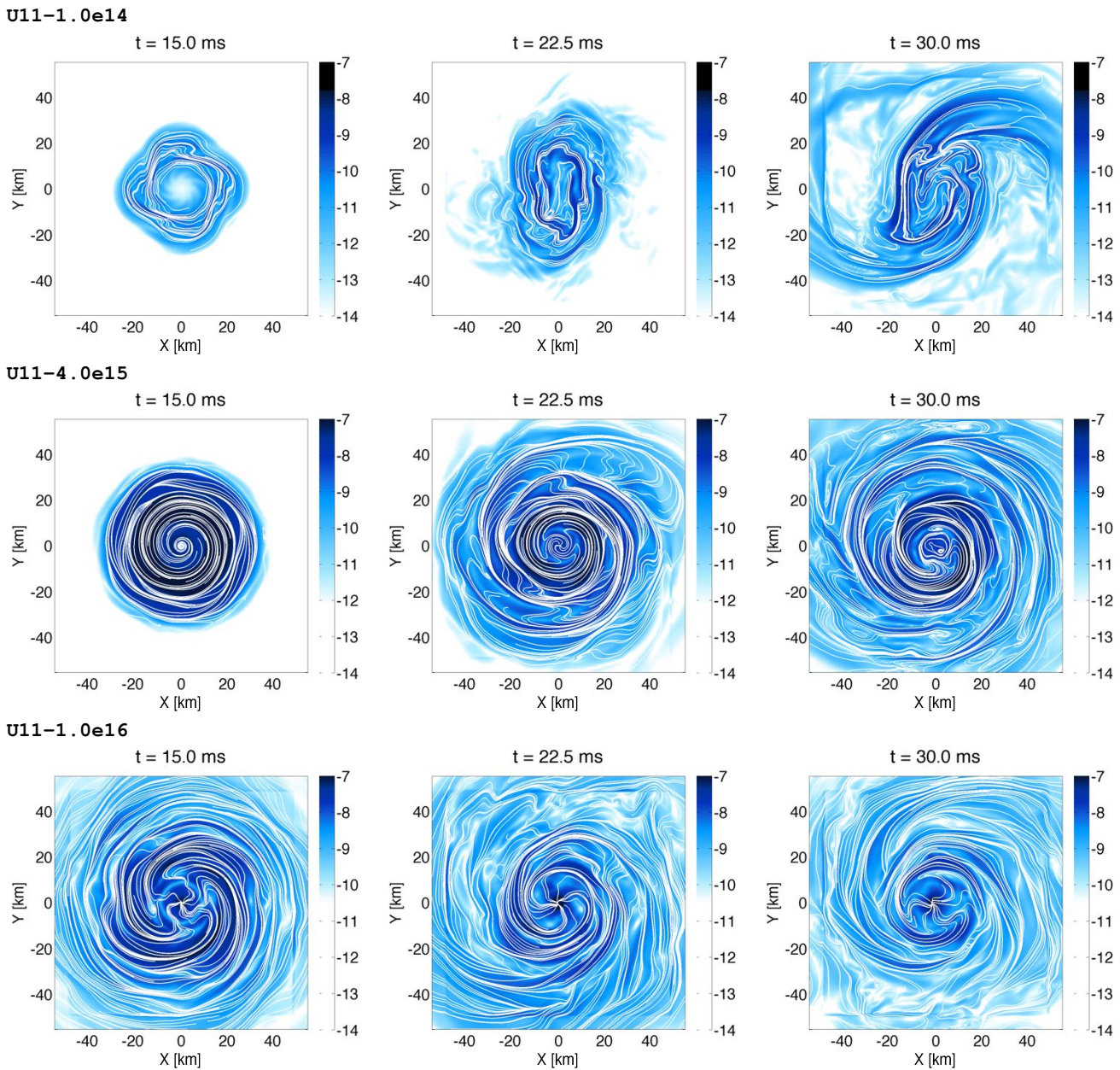

\noindent
\begin{tabular}{ccc}
\multicolumn{3}{l}{\bf \texttt{U11-1.0e14}}\\
          \includegraphics[width=0.32\textwidth]{\imgname{U11B1e14_EMt00_151}}
&         \includegraphics[width=0.32\textwidth]{\imgname{U11B1e14_EMt00_225}}
&         \includegraphics[width=0.32\textwidth]{\imgname{U11B1e14_EMt00_300}}
\\
\multicolumn{3}{l}{\bf \texttt{U11-4.0e15}}\\
          \includegraphics[width=0.32\textwidth]{\imgname{U11B4e15_EMt00_151}}
&         \includegraphics[width=0.32\textwidth]{\imgname{U11B4e15_EMt00_225}}
&         \includegraphics[width=0.32\textwidth]{\imgname{U11B4e15_EMt00_300}}
\\
\multicolumn{3}{l}{\bf \texttt{U11-1.0e16}}\\
          \includegraphics[width=0.32\textwidth]{\imgname{U11B1e16_EMt00_151}}
&         \includegraphics[width=0.32\textwidth]{\imgname{U11B1e16_EMt00_225}}
&         \includegraphics[width=0.32\textwidth]{\imgname{U11B1e16_EMt00_300}}
\\
\end{tabular}
\caption{Snapshots of the total electromagnetic energy density
  $T^{00}_{\textrm{em}}$, as measured in the Eulerian frame, on a
  horizontal plane at $z\simeq 1.5$ km for model \texttt{U11-1.0e14}
  (top row), \texttt{U11-4.0e15} (central row), and
  \texttt{U11-1.0e16} (right row), at different times during the
  evolution, namely, $t = 15.0$ ms (left column), $t=22.5$ ms (central
  column), and $t=30$ ms (right column).  The magnetic field lines are
  shown with white solid lines.}%
   \label{fig:EmagSnapshots}
\end{figure*}

We complete the description of the dynamics of these three
\texttt{U11} models by reporting in the right column of
Fig.~\ref{fig:spacetime_U11} the spacetime diagram relative to the
angular velocity $\Omega$ along the $x$-axis. We recall that all
models have the maximum of the $\Omega$ at the stellar center (\cf
Fig.~\ref{fig:InitialModels}) and this remains the case also for the
low-magnetic-field and bar-mode unstable model \texttt{U11-1.0e14},
modulo the variations brought in by the development of the
instability. On the other hand, for models \texttt{U11-4.0e15} and
\texttt{U11-1.0e16}, the angular velocity at the stellar center first
increases, then reaches a maximum and later decreases again; at the
same time, the outer layers of the star expand and the maximum of the
angular velocity occurs at larger radii. By the time an extended
flattened envelope has been produced near the equatorial plane, much
of the differential rotation has been washed out and the NS has
acquired a central angular velocity that is smaller but mostly uniform.

We can summarize the main features described in
detail above for the three magnetized \texttt{U11} models as follows:
\begin{itemize}
\item model \texttt{U11-1.0e14} is still bar-mode unstable and no
  effects are evident on the onset and development of the instability;
  a very small fraction of the rest-mass is shed at the edges of the
  bar-deformed object.
  \item model \texttt{U11-4.0e15} is bar-mode stable for the
    timescales considered here and after about 25 ms of evolution it
    settles into a more compact configuration; the new equilibrium
    structure has an almost uniform angular velocity and is surrounded
    by a differentially and flattened envelope.
  \item model \texttt{U11-1.0e16} is also bar-mode stable with a
    dynamics that resembles that of model \texttt{U11-4.0e15}; the
    main differences are the shorter timescales required to reach
    equilibrium and the flattened envelope with larger mean rest-mass
    densities present in model \texttt{U11-4.0e15}.
\end{itemize}
Altogether, the behavior summarized above is consistent with what we
would expect for highly magnetized and differentially rotating
fluids. Under these conditions, in fact, magnetic braking transfers
angular momentum from the core to the outer layers, changing the
rest-mass density and the rotation profiles of the star. Because
during this process part of the rotational energy of the star is
tapped, the onset of the instability is inhibited.

We next discuss the dynamics of the magnetic fields, in
Fig.~\ref{fig:EmagSnapshots}, and show for comparison representative
snapshots of the total electromagnetic energy density
$T^{00}_{\textrm{em}}$ (shown with a colorcode) as measure in the
Eulerian frame and the magnetic field lines (shown as white solid lines),
as measured on a horizontal plane at $z\simeq 1.5$ km, corresponding to
the three magnetized \texttt{U11} models studied before. Note that all
the panels have the same color ranges but that the colormap is different
at different times (\ie in different columns) in order to better
highlight the internal structure of the electromagnetic field. The
various columns refer to different times and coincide with those already
reported in Fig.~\ref{fig:RhoSnaphots}.

As expected under the ideal-MHD approximation, with the magnetic field
being frozen into the fluid, the field lines are dragged along with
the fluid in differential rotation and rapidly wind on a timescale of
very few milliseconds, leading to a sudden formation and rapid linear
growth of a toroidal magnetic field component. This component is soon
amplified far above the initial poloidal one. The winding of the field
lines and the linear growth of the toroidal field are present in all
three models and are independent of the initial poloidal magnetic
field strength. The reason is that they only depend on the angular
velocity profile, or equivalently on the differential rotation law,
which is the same for all \texttt{U11} models in the first part of the
evolution. It interesting to note in the first row of the figure (\ie
the unstable model \texttt{U11-1.0e14}), that the distortion of the
magnetic field lines also mimics the bar-mode deformation as the star
undergoes the development of the instability.

\begin{figure}
\begin{center}
\includegraphics[width=0.48\textwidth]{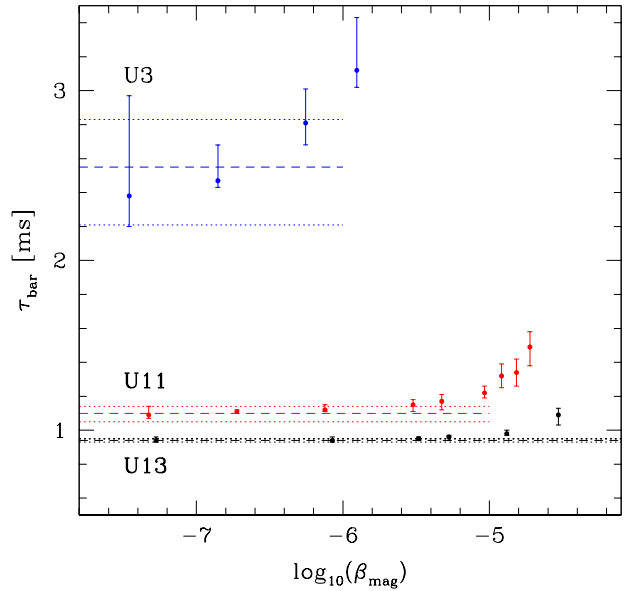}
\end{center}
\caption{Growth time of the bar-mode instability for the three unstable
  models \texttt{U3} (blue), \texttt{U11} (red) and \texttt{U13} (black),
  shown as a function of the initial magnetization. The horizontal dashed
  lines report the growth times in the absence of magnetic fields, while
  the dotted lines represent the corresponding error bars.}
\label{fig:Growth_time}
\end{figure}

A more quantitative assessment of the influence of the magnetic fields on
the unstable models has been obtained after performing a number of
simulations of models \texttt{U3}, \texttt{U11} and \texttt{U13}, with
initial poloidal magnetic fields varying between the two extreme cases
presented in Figs.~\ref{fig:RhoSnaphots}--\ref{fig:EmagSnapshots}. More
specifically, we have performed 27 simulations with initial maximum
magnetic fields in the range $B^{z}_\text{max}|_{t,z=0} = 1.0 \times
10^{14}$ and $1.0 \times 10^{16}$ G. The results of this extensive
investigation are collected in Figs.~\ref{fig:InitialModels}
and~\ref{fig:Growth_time}, as well as in
Table~\ref{table:barpropertiesU11}, which reports the measured growth
time of the instability $\tau_{\text{bar}}$ and its frequency
$f_{\textrm{bar}}$. In particular, Fig.~\ref{fig:InitialModels} reports
the initial models within a $(\beta,\,\beta_{\text{mag}})$ diagram and
allows one to easily distinguish the ranges of rotational and magnetic
energies that give rise to the development of a dynamical bar-mode
instability. It is, in fact, easy to distinguish models that are bar-mode
stable (blue symbols) from those that are unstable (red symbols) at zero
magnetizations; of course, models that are stable at zero magnetizations
are also stable at all magnetizations (this is marked with the vertical
red dashed line). Equally simple is to distinguish models that although
unstable in the absence of magnetic fields (red squares), become stable
with sufficient magnetization (red triangles). As an example, for models
\texttt{U3} the threshold between squares and triangles appears for
initial maximum magnetic fields $B^{z}_\text{max}|_{t,z=0} > 6.0 \times
10^{14}$ G, while for models \texttt{U11} and \texttt{U13} the threshold
is at about $2.0 \times 10^{15}$ and $2.4 \times 10^{15}$ G
respectively. As a result, only the light-red shaded area in
Fig.~\ref{fig:InitialModels} collects stellar models that are bar-mode
unstable. Outside this region, either the rotational energy is
insufficient, or the magnetic tension is too strong to allow for the
development of the instability.

Similarly, Fig.~\ref{fig:Growth_time} reports the measured growth time
of the instability $\tau_{\text{bar}}$ (and the corresponding error
bars) for the three different classes of unstable models (\texttt{U3},
\texttt{U11} and \texttt{U13}) as a function of the magnetization
parameter $\beta_{\text{bar}}$. Taking the horizontal dashed lines as
references for the unmagnetized models, it is easy to realize that as
the magnetization increases, so does the growth time for the
instability. This behavior can be physically interpreted as due to the
fact that as the magnetic field strength increases, so does the
timescale over which the magnetic tension needs to be won to develop a
bar deformation\footnote{Note that the error bars are larger for model
  \texttt{U3} because this is closer to the stability threshold (\cf
  Table~\ref{table:models}).}.

\begin{figure*}
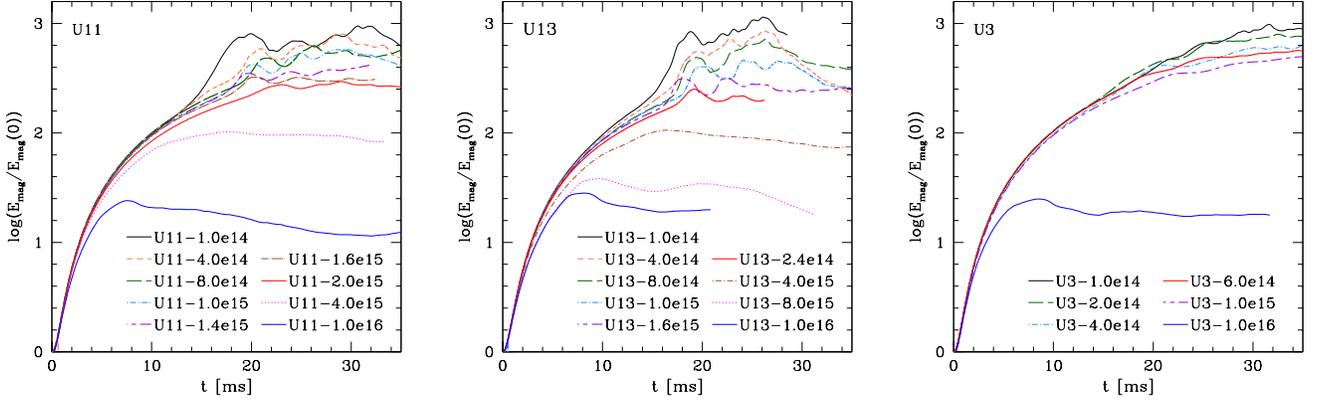

\begin{center}
\includegraphics[width=0.33\textwidth]{\imgname{betamag_U11}}
\includegraphics[width=0.33\textwidth]{\imgname{betamag_U13}}
\includegraphics[width=0.33\textwidth]{\imgname{betamag_U03}}
\end{center}
\caption{\textit{Left panel:} evolution of the total magnetic energy
  $E_{\text{mag}}$, normalized to its initial value, for models
  \texttt{U11} and different values of the initial poloidal magnetic
  field. The black solid line refers to the less magnetized case, the
  blue solid line for the most magnetized case, and a red solid line for
  the last unstable model, before excessive magnetic tension suppresses
  the instability. \textit{Middle and right panels:} the same as in the
  left panel, but for model \texttt{U13} and model \texttt{U3},
  respectively.}
\label{fig:EmagU11}
\end{figure*}

We can next focus on the growth of the magnetic-field strength in
bar-mode unstable models as this also offers the opportunity for a
number of useful considerations. More specifically, we show in
Fig.~\ref{fig:EmagU11} the evolution of the total electromagnetic
energy $E_{\text{mag}}$ normalized to the initial values for models
\texttt{U11} (left panel), \texttt{U13} (middle panel) and \texttt{U3}
(right panel), and for different initial poloidal magnetic-field
strengths. The first obvious thing to notice in Fig.~\ref{fig:EmagU11}
for all the magnetizations considered for model \texttt{U11} is that
the growth of the magnetic energy is linear in time initially. This is
not surprising and is indeed the mere manifestation of the
``frozen-in'' condition of the magnetic field within the ideal-MHD
approximation. Using the induction equation it is, in fact,
straightforward to show that in a linear regime the differential
rotation will generate toroidal magnetic field at a rate which is
linear in time. This is because as long as the stellar configuration
remains essentially axisymmetric the poloidal magnetic field is not
affected by the newly produced toroidal field, and the total
electromagnetic energy can only grow linearly with time tapping part
of the rotational energy of the star.

As a result of this growth, the toroidal component becomes rapidly
larger than the initial poloidal one and an amplification of the
total electromagnetic energy takes place for all models that reaches a
higher value of about two orders of magnitude over a timescale of
$\sim 10$ ms. After this initial phase, the toroidal field keeps
growing at a slower rate, reaching a saturation with the maximum
amplification being almost independent of the initial poloidal
magnetic field strength and of the rotation of the stellar model. The
only exceptions to this behavior appear in models with ultra-strong
magnetic fields, in which cases the saturation occurs at values that
are about two orders of magnitude smaller (\cf blue solid lines in the
different panels of Fig.~\ref{fig:EmagU11}).

\begin{table}
\begin{tabular}{|c|c|cc|c|c|c|}
\hline\hline 
 Model & $\beta_\text{mag}$ & $t_1$ & $t_2$ & $\eta_{\text{max}}$ & $\tau_{\text{bar}}$ & $f_{\textrm{bar}}$  \\
       &                    & [ms]  & [ms]  &                     &   [ms]              &  [Hz] \\
\hline   
\texttt{U11-0.0e00}  & $   0.0          $  & $16.2$ & $18.3$ & $0.784$ & $1.10^{+0.04}_{-0.05}$ & $490^{+1}_{-4}$ \\ 
\texttt{U11-1.0e14}  & $4.7\times 10^{-8}$  & $14.7$ & $16.8$ & $0.787$ & $1.09^{+0.05}_{-0.02}$ & $491^{+3}_{-5}$ \\
\texttt{U11-2.0e14}  & $1.9\times 10^{-7}$  & $15.0$ & $17.0$ & $0.778$ & $1.11^{+0.02}_{-0.01}$ & $488^{+1}_{-1}$ \\
\texttt{U11-4.0e14}  & $7.5\times 10^{-7}$  & $15.1$ & $17.7$ & $0.773$ & $1.12^{+0.03}_{-0.01}$ & $488^{+2}_{-2}$ \\
\texttt{U11-8.0e14}  & $3.0\times 10^{-6}$  & $14.8$ & $18.2$ & $0.754$ & $1.15^{+0.03}_{-0.04}$ & $490^{+2}_{-5}$ \\
\texttt{U11-1.0e15}  & $4.7\times 10^{-6}$  & $14.2$ & $16.8$ & $0.751$ & $1.17^{+0.04}_{-0.05}$ & $491^{+2}_{-4}$ \\
\texttt{U11-1.4e15}  & $9.2\times 10^{-6}$  & $13.9$ & $16.2$ & $0.714$ & $1.22^{+0.04}_{-0.03}$ & $491^{+1}_{-2}$ \\
\texttt{U11-1.6e15}  & $1.2\times 10^{-5}$  & $14.5$ & $17.3$ & $0.681$ & $1.32^{+0.07}_{-0.07}$ & $489^{+2}_{-1}$ \\
\texttt{U11-1.8e15}  & $1.5\times 10^{-5}$  & $13.2$ & $16.7$ & $0.639$ & $1.34^{+0.08}_{-0.08}$ & $490^{+2}_{-1}$ \\
\texttt{U11-2.0e15}  & $1.9\times 10^{-5}$  & $14.8$ & $17.3$ & $0.532$ & $1.49^{+0.09}_{-0.11}$ & $489^{+4}_{-2}$ \\
\hline    
\texttt{U13-0.0e00}  & $   0.0          $  &  $11.6$ & $14.7$ & $0.865$ & $0.94^{+0.01}_{-0.01}$ & $449^{+1}_{-3}$ \\
\texttt{U13-1.0e14}  & $5.3\times 10^{-8}$  &  $12.2$ & $15.3$ & $0.866$ & $0.94^{+0.02}_{-0.01}$ & $450^{+2}_{-2}$ \\
\texttt{U13-4.0e14}  & $8.5\times 10^{-7}$  &  $12.7$ & $15.8$ & $0.851$ & $0.94^{+0.02}_{-0.01}$ & $450^{+2}_{-2}$ \\
\texttt{U13-8.0e14}  & $3.3\times 10^{-6}$  &  $12.7$ & $15.8$ & $0.842$ & $0.95^{+0.01}_{-0.01}$ & $451^{+1}_{-2}$ \\
\texttt{U13-1.0e15}  & $5.3\times 10^{-6}$  &  $14.1$ & $16.7$ & $0.833$ & $0.96^{+0.01}_{-0.02}$ & $451^{+3}_{-1}$ \\
\texttt{U13-1.6e15}  & $1.3\times 10^{-5}$  &  $11.6$ & $14.8$ & $0.813$ & $0.98^{+0.02}_{-0.01}$ & $456^{+1}_{-2}$ \\
\texttt{U13-2.4e15}  & $3.0\times 10^{-5}$  &  $13.0$ & $15.9$ & $0.734$ & $1.09^{+0.04}_{-0.06}$ & $461^{+1}_{-1}$ \\
\hline
\texttt{U3-0.0e00}   & $   0.0          $  &  $24.8$ & $26.4$ & $0.486$ & $2.55^{+0.28}_{-0.34}$ & $540^{+2}_{-2}$ \\
\texttt{U3-1.0e14}   & $3.5\times 10^{-8}$  &  $24.9$ & $27.1$ & $0.472$ & $2.38^{+0.59}_{-0.18}$ & $537^{+5}_{-10}$ \\
\texttt{U3-2.0e14}   & $1.4\times 10^{-7}$  &  $26.1$ & $28.0$ & $0.456$ & $2.47^{+0.21}_{-0.04}$ & $536^{+5}_{-3}$ \\
\texttt{U3-4.0e14}   & $5.6\times 10^{-7}$  &  $24.0$ & $26.3$ & $0.421$ & $2.81^{+0.20}_{-0.13}$ & $537^{+2}_{-3}$ \\
\texttt{U3-6.0e14}   & $1.2\times 10^{-6}$  &  $24.2$ & $25.7$ & $0.300$ & $3.12^{+0.31}_{-0.10}$ & $535^{+5}_{-6}$ \\
\hline
\hline
\end{tabular}
\caption{Main properties of the initial part of the instability for
  model \texttt{U11}, \texttt{U13} and \texttt{U3} for different
  values of the initial poloidal magnetic field. Here we report the
  representative times $t_1$ and $t_2$ between which the maximum values of the
  distortion parameter $\eta$, the growth times $\tau_{\text{bar}}$
  and the frequencies $f_{\textrm{bar}}$ are computed.}
\label{table:barpropertiesU11}
\end{table}

Interestingly, for models \texttt{U11} and \texttt{U13}, that is for the
unstable models with small growth rates and far from the threshold of the
dynamical bar-mode instability, the linear growth of the magnetic field
is accompanied also by a rather short exponential growth of the magnetic
field. While this behavior is very similar to the one seen in
Ref.~\cite{Siegel2013}, where it was attributed to the development of the
MRI, a similar conclusion cannot be drawn with confidence here. On the
one hand, there are a number of combined elements that seem to support
the suggestion that the exponential growth is the result of the
development of an MRI: (i) the instability disappears with decreasing
resolution (the smallest wavelength needs to be properly resolved); (ii)
the growth rate does not depend on the initial poloidal magnetic field
(in the simplest description the growth rate depends only on the local
angular velocity); (iii) the exponential growth is followed by a rapid
decay possibly caused by reconnection processes (this behavior was also
found in Ref.~\cite{Siegel2013}); (iv) the exponential growth disappears
for sufficiently strong magnetic fields (the bar-mode deformation is no
longer the lowest energy state energetically because of the large
magnetic-field contribution). However, our resolutions here are
considerably coarser than those employed in Ref.~\cite{Siegel2013}, and
it is therefore difficult to see the appearance of channel-flow
structures typical of the MRI and hence to make robust measurements of
the wavelengths of the fastest-growing modes. One important feature of
models \texttt{U11} and \texttt{U13} is that they develop pronounced
bar-mode deformations (they are further away from the stability threshold
in Fig.~\ref{fig:InitialModels}) and it is therefore possible that these
large deviations from axisymmetry act as an additional trigger, favouring
the development of the MRI\footnote{We recall that the assumption of
  axisymmetry is a fundamental one in all perturbative calculations on
  the MRI and that it is exactly the absence of axisymmetry that allows
  for the development of dynamos against the limitations of the Cowling
  theorem~\cite{Choudhuri1998}.}. This could explain why an exponential
growth is seen in these models despite the coarse resolution. At the
moment this is just a conjecture, which however, if confirmed, could shed
light on the sufficient conditions for the development of the MRI and in
particular on the degree of axisymmetry needed by the system. Additional
simulations at much higher resolutions will be necessary to address this
point in the future.

Interestingly, no exponential growth has been measured in the dynamics of
model \texttt{U3} for all the different magnetizations considered (\cf
right panel of Fig.~\ref{fig:EmagU11}). Although the angular frequency of
these models is larger than that of \texttt{U11} and \texttt{U13} and
hence the timescale for the development of the MRI $\tau_{_\textrm{MRI}}$
would be correspondingly shorter ($\tau_{_\textrm{MRI}} \sim
\Omega^{-1}$). The evolutions have been carried out on sufficiently long
timescales to allow for the potential appearance of the MRI. This
behavior is indeed consistent with the conjecture discussed above,
since this class of models is very close to the threshold for the
development of the bar-mode instability. As a result, these models
experience much smaller bar-mode deformations and maintain a
configuration which is more axisymmetric than those found in models
\texttt{U11} and \texttt{U13}. Because these conditions are more
similar to those assumed by perturbative MRI analysis, the
corresponding predictions are expected to be more accurate. Hence, it
is not surprising that no MRI is observed in this case simply because
no MRI can be seen for these quasi-axisymmetric objects at these
resolutions.

\subsection{Effects of the magnetic field on stable models}
\label{subsec:StableModels}

\begin{figure}
\begin{center}
\includegraphics[width=0.48\textwidth]{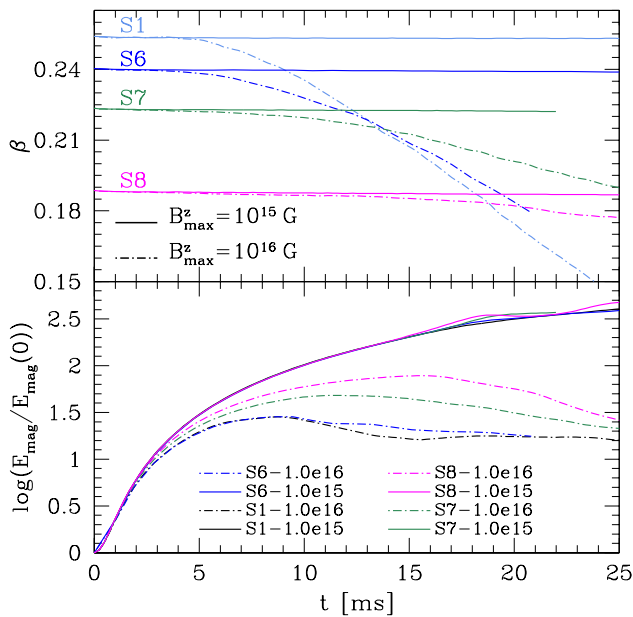}
\end{center}
\caption{Evolution of the rotation parameter $\beta := T / |W|$ (top
  panel) and of the total magnetic energy normalized to its initial value
  (bottom panel) for models \texttt{S1}, \texttt{S6}, \texttt{S7} and
  \texttt{S8} which are stable against the bar-mode deformation in the
  un-magnetized case. In both panels the solid lines refer to models with
  $B^{z}_\text{max}|_{t,z=0}=10^{15}$ G, while the dash-dot lines to
  models with $B^{z}_\text{max}|_{t,z=0}=10^{16}$ G.}
\label{fig:dynamics_stable_models}
\end{figure}

After having discussed in detail the properties of the dynamics of
bar-mode \emph{unstable} models, we now turn to illustrating how magnetic
fields affect the dynamics of bar-mode \emph{stable} models. Although
these are comparatively simpler configurations, they provide a number of
interesting considerations, as we will see. 

We recall that using the same EOS adopted here, Ref.~\cite{Baiotti06b}
has determined the threshold for the development of a dynamical bar-mode
instability to be $\beta \simeq 0.255$ (\cf
Fig.~\ref{fig:InitialModels}). We have therefore considered a number of
stable models, namely \texttt{S1}, \texttt{S6}, \texttt{S7} and
\texttt{S8}, that are increasingly more distant from the threshold. For each
of these classes we have then added two different magnetic-field
strengths, namely, $B^{z}_\text{max}|_{t,z=0}=1.0 \times 10^{15}$ G and
$B^{z}_\text{max}|_{t,z=0}=1.0 \times 10^{16}$ G, and performed
simulations to record the different impact of the magnetic fields on the
dynamics.

Of course, since these models are already stable in the absence of
magnetic fields, they will remain stable also with the additional
magnetic tension. However, while models with
$B^{z}_\text{max}|_{t,z=0}=1.0 \times 10^{15}$ G do not show in their
dynamics any significant deviation from a purely hydrodynamical
evolution, models with $B^{z}_\text{max}|_{t,z=0}=1.0 \times 10^{16}$
G do quite the opposite. This is shown in the top panel of
Fig. \ref{fig:dynamics_stable_models}, which reports the evolution of
the rotation parameter $\beta$ for all these stable models. Solid
lines of different color refer to the different models but all having
an initial magnetic field $B^{z}_\text{max}|_{t,z=0}=1.0 \times
10^{15}$ G. On the other hand, dot-dashed lines of different color
refer to models with $B^{z}_\text{max}|_{t,z=0}=1.0 \times 10^{16}$
G. Note that for comparatively ``low'' magnetic fields, the rotation
parameter does not show any significant variation from the initial
value over a timescale of around $25$ ms, with changes that are
$\lesssim 0.4\%$ for model \texttt{S1} and $\lesssim 1.0\%$ for model
\texttt{S8}. On the other hand, for magnetic fields that are one order
of magnitude larger, the rotation parameter changes significantly,
decaying almost linearly with time. This is obviously due to the
combined action of the differential rotation and of the magnetic
winding, which increases the magnetic tension and drives the NS
towards a configuration that is uniformly rotating. This is also
very clearly shown in the bottom panel of
Fig. \ref{fig:dynamics_stable_models}, which reports the evolution of
the normalized magnetic energy. It is then rather clear that while the
energy increases (linearly) with time in the case of comparatively
small magnetic fields (solid lines), it stops growing and saturates in
the case of large magnetic fields (dot-dashed lines). Over the
timescale of the simulations, $\sim 25$ ms, the magnetic energy has
increased of almost three orders of magnitude in the former case and
of only one in the latter case.

We can use the results in the top panel of
Fig. \ref{fig:dynamics_stable_models} to obtain an important estimate on
the rate at which the stellar rotational energy is completely tapped by
the generation of a toroidal magnetic field. In particular, using the
numerical data it is possible to express $\beta$ as
\begin{equation}
\beta(t) \simeq \beta_0 + a \exp\left(1-\frac{b}{t}\right) - c t\,,
\label{eq:betafit}
\end{equation}
where $\beta_0 := \beta(t=0)$ and $a,b, c$ are three constant
coefficients to be computed from a fit to the numerical data.  Using
this expression is possible to compute the ``braking timescale''
$\tau_{\textrm{br}}$, that is, the timescale needed for an
axisymmetric and differentially rotating configuration to lose all of
its rotational energy via magnetic-field shearing and thus be brought
to have $\beta(\tau_{\textrm{br}}) = 0$\footnote{It is perfectly
  plausible that a nonrotating configuration is never reached because
  the system will first collapse to a black hole. Of course this is
  possible only for initial stellar configurations that are
  supermassive (see~\cite{Falcke2013} for the exploration of this
  possibility).}.  The numerical fits show that the parameter $b$ we
obtain is of the order of the simulation time and indeed the
expression in Eq.\ (\ref{eq:betafit}) is used only for time
$t<b$. Considering the limit $t \gg b$ in the expression above and
after a little bit of algebra we can show that the timescale is
\begin{equation}
\label{eq:braking_time}
\tau_{\textrm{br}} \simeq \frac{\beta_0 + a e}{c}\,.
\end{equation}
Interestingly, for the EOS and the magnetic field of $10^{16}$ G
considered here, the coefficients $a$ and $c$ have a simple dependence on
$\beta_0$\footnote{The coefficient $b$ describes the transient and is not
relevant to study the asymptotic solution.}, \ie
\begin{align}
\label{eq:braking_time_coeffs}
a &\simeq  k_1 + k_2 \beta_0 \,,& \qquad c &\simeq k_3 + k_4 \beta_0\,,
\end{align}
where $k_1 \simeq -0.0589$, $k_2 \simeq 0.332$, $k_3 \simeq -0.017$
ms${}^{-1}$ and $k_4 \simeq 0.092$
ms${}^{-1}$. As a result, the general expression for the braking time is now just
a function of the initial rotation parameter
\begin{equation}
\label{eq:braking_time_new}
\tau_{\textrm{br}} \simeq 
\frac{\left[ \beta_0 + e\, \left(k_1 + k_2 \beta_0\right)\right]}{k_3 + k_4 \beta_0}\,,
\end{equation}
and of the four coefficients $k_1, k_2, k_3$ and $k_4$. As an example, we can
use the expression~\eqref{eq:braking_time_new} to estimate the braking time
for models \texttt{S1} and \texttt{S8}, readily obtaining
$\tau_{\textrm{br}} \sim 0.050$ s and $\tau_{\textrm{br}} \sim 0.56$ s,
respectively.

\begin{figure}
\begin{center}
\includegraphics[width=0.48\textwidth]{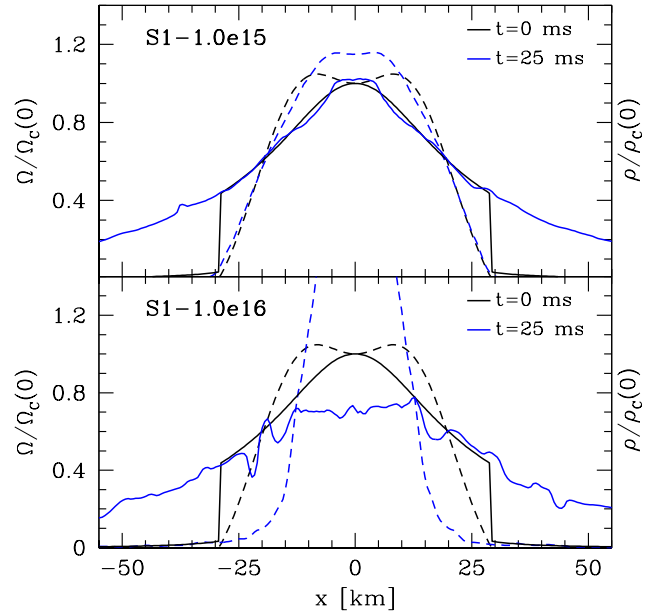}
\end{center}
\caption{Initial (\ie shown with black lines at $t=0$ ms) and the final
  (\ie shown with blue lines at $t=25$ ms) normalized profiles along the
  $x$-direction of the angular velocity (solid lines) and of the
  rest-mass density (dashed lines) for model \texttt{S1} and the two
  different values of the magnetic field, \ie $10^{15}$ G (top panel) and
  $10^{16}$ G (bottom panel).}
\label{fig:Omega_profile_stable}
\end{figure}

Much of what is discussed above can also be deduced when analyzing the
structural changes in the NSs. These are shown in
Fig.~\ref{fig:Omega_profile_stable}, where we report the initial (\ie at
$t=0$ ms with black lines) and the final (\ie at $t=25$ ms with blue
lines) normalized profiles along the $x$-direction of the angular
velocity (solid lines; \cf left $y$-axis of the figure) and of the
rest-mass density (dashed lines; \cf right $y$-axis of the figure) for
model \texttt{S1} and the two different values of the magnetic field
($B^{z}_\text{max}|_{t,z=0}=1.0 \times 10^{15}$ G in the top panel and
$B^{z}_\text{max}|_{t,z=0}=1.0 \times 10^{16}$ G in the bottom one).

Modulo of course the fact that the star will have shed some matter and
produced a more extended, very low-density outer mantle, it is
clear from the top panel of Fig.~\ref{fig:Omega_profile_stable} that
the angular velocity and rest-mass density in the stellar core hardly
change. This is to be contrasted with what shown in the bottom panel,
which clearly shows a very large increase of the rest-mass density in
the inner regions of the star and a corresponding decrease in the
outer ones. At the same time, the angular velocity profile has
flattened considerably and indeed the star is essentially axisymmetric
and in uniform rotation within a coordinate radius of $\simeq 15$ km.

As a final remark, we note that the considerations made here and in
particular the braking timescale estimated in
Eq.~\eqref{eq:braking_time}, are of course of more general validity than
the bar-mode instability context considered here. Compact, axisymmetric
and differentially rotating magnetized configurations can in fact be
produced in a number of scenarios, from stellar-core collapse to the
merger of binary NSs. Determining the timescales over which uniform
rotation is established under controlled setups in terms of
differential-rotation laws and magnetizations is of course of great
importance and will be addressed in a future work.

\section{Conclusions}
\label{sec:conclusions}

We have presented a study of the dynamical bar-mode instability in
differentially rotating and magnetized NSs in full general relativity and
investigated how the presence of magnetic fields affects the onset and
the development of the instability. In order to do that, we have
performed 3D ideal-MHD simulations of a large number of stellar models
that were already studied in the absence of magnetic
fields~\cite{Baiotti06b,Manca07,DePietri06}, by adding an initial purely
poloidal magnetic field with strengths between $10^{14}$ and $10^{16}$ G.
In this way, we were able to explore quite extensively the parameter
space $(\beta, \, \beta_{\rm{mag}})$ from $\beta=0.1886$ to $0.2812$,
determining a threshold for the onset of the instability both in terms of
the rotation parameter $\beta = T/|W|$ and of the magnetization parameter
$\beta_{\rm{mag}}=E_{\text{mag}}/(T+|W|)$. In all cases considered, the
differential rotation shears the poloidal magnetic field, generating a
toroidal component that grows linearly in time, and which soon provides
the largest contribution to the total electromagnetic energy.

When considering initial stellar models that are bar-mode unstable in the
absence of magnetic fields, we found that no effects are present on the
dynamics of the bar-mode deformation for initial poloidal magnetic fields
that are $\lesssim 10^{15}$ G, with the exact threshold depending on the
rotational properties of the initial model and being higher for slower
rotating models. This is not particularly surprising given that in these
cases the magnetic energy, even the one produced via magnetic-field
shearing, is only a small contribution to the total energy of the
system. For initial magnetic fields that are instead $\gtrsim 10^{16}$ G
or larger, the corrections introduced by the magnetic tension become
quite large. In particular, below a critical $\beta_{\rm{mag}}$, the
development of the instability is modified, showing growth rates and
bar-mode distortions that become smaller with increasing magnetic fields,
and possibly exhibiting an exponential growth of the toroidal component
at later times. Above a critical $\beta_{\rm{mag}}$, on the other hand,
the instability is totally suppressed as the enormous magnetic tension
cannot be overcome by the differential rotation. Under these conditions,
the star sheds its outer layers leading to an extended, axisymmetric
object with a high, uniform-density core and a low-density, slowly
rotating envelope.

On the basis of the phenomenology discussed above, and after carrying-out
a large number of simulations, we were able to locate in the $(\beta, \,
\beta_{\rm{mag}})$ diagram the regions in which the values of the
rotational and magnetic energies are sufficient to give rise to the
development a dynamical bar-mode instability. In this sense, our study
confirms the Newtonian results of~\cite{Camarda:2009mk} and extends them
to a general-relativistic framework and to a more generic range of
initial conditions.

We have complemented our investigation by considering also initial
stellar models that are bar-mode stable in the absence of magnetic
fields. While these are comparatively simpler configurations, also in
this case the magnetic fields can provide structural changes if
sufficiently strong. More specifically, for magnetic fields $\gtrsim
10^{16}$ G, the stellar models are braked considerably in their rotation
and evolve into configurations that have uniformly rotating extended
cores with large rest-mass densities when compared to the initial
values. A simple algebraic expression has been derived to estimate the
timescale over which an axisymmetric and differentially rotating
configuration will lose all of its rotational energy via magnetic-field
shearing.

As a final remark we note that although we have restricted our attention
to a simplified EOS, our results also point out that it is unlikely that
very highly magnetized NSs can develop the dynamical bar-mode instability
and hence be considered as strong sources of GWs.

\begin{acknowledgments}

  We have benefited from discussions with several colleagues and friends
  and we are particularly grateful to Alessandra Feo, Sebastiano
  Bernuzzi, Niccol\'o Bucciantini, Riccardo Ciolfi, Luca Del Zanna,
  Filippo Galeazzi, and Frank L\"offler. This work has been partially
  supported by the ``CompStar'', a Research Networking Programme of the
  European Science Foundation and the DFG grant SFB/Transregio 7. The
  calculations have been performed using the HPC resources of the INFN
  ``Theophys'' cluster, the PRACE allocation (6th-call) ``3dMagRoI'' on
  CINECA's supercomputer ``Fermi'', the cluster `at the Albert-Einstein
  Institute, and the PRACE allocation ``pr32pi'' on ``SuperMUC'' at the
  ``Leibniz-Rechenzentrum''.
  
\end{acknowledgments}

\appendix

\section{The role of symmetries}
\label{app_a}

\begin{figure}
\begin{center}
\includegraphics[width=0.49\textwidth]{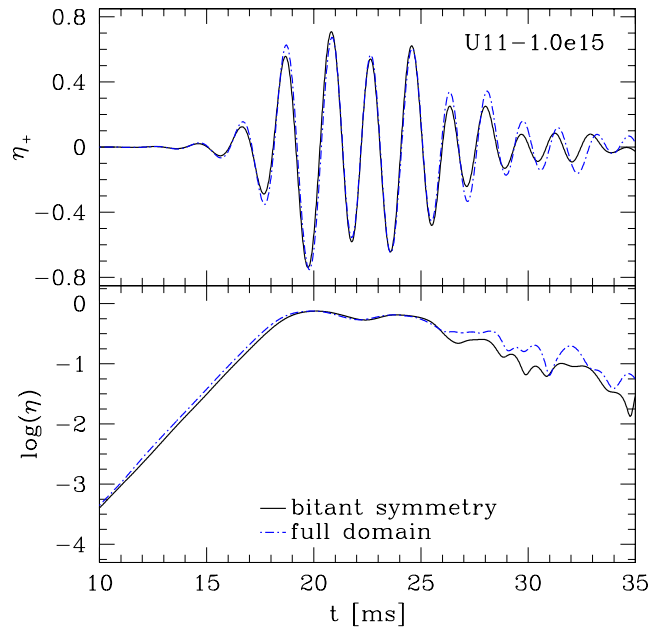}
\end{center}
\caption{Evolution of the distortion parameters $\eta_+$ (top panel) and
  $\eta$ (bottom panel) for model \texttt{U11-1.0e15} when imposing a
  bitant symmetry (black solid line) or when using the full domain (blue
  dashed line).}
\label{fig:symmetry}
\end{figure}

As discussed in Sect.~\ref{sec:evolution}, all of the results presented
here were achieved with a spatial resolution $\Delta x = 0.375 \, M_\odot
\simeq 0.550$ km on the finest grid and exploiting a ``bitant symmetry'',
\ie a reflection symmetry with respect to the $(x,y)$ plane. While this
choice obviously reduces the computational costs by a factor two, it is
important to verify that it does not introduce systematic effects and
that all the results would be unchanged if this symmetry was suppressed. 

Although the \texttt{WhyskyMHD} code employed here has been tested in a
number of different scenarios and its accuracy has already been
explicitly reported in various works \cite{Giacomazzo:2007ti,
  Giacomazzo:2010, Rezzolla:2011}; nevertheless, we have performed
additional tests to check that the specific settings used are sufficient
to capture the main properties of the evolved systems. To this scope we
have evolved the bar-mode unstable model \texttt{U11} when threaded by an
initially moderate magnetic field, \ie model \texttt{U11-1.0e15}, both
when imposing the bitant symmetry and when evolving the equations in the
full domain. Furthermore, we have varied the resolution of more than a
factor of two, that is, with the finest grid having resolutions between
$\Delta x=0.370$ km and $0.920$ km.

For all these runs we computed the growth rate, $\tau_{\rm{bar}}$, and
the frequency, $f_{\textrm{bar}}$, of the bar-mode instability. The
results of this extensive series of tests are reported in
Table~\ref{table:resolution}) and show that these quantities do not
depend on resolution within the accuracy of our estimate. Hence, we
conclude that all of the results have been achieved at sufficient
resolution to extract physically significant information.

In addition, we have also verified that no systematic effects have
been introduced by the use of a bitant symmetry and this is shown in
Fig.~\ref{fig:symmetry}, where we report the evolution of the
distortion parameters $\eta_+$ (top panel) and $\eta$ (bottom panel)
for model \texttt{U11-1.0e15}. The simulations have been performed at
the reference resolution of $\Delta x = 0.550$ km on the finest grid,
and the figure offers a comparison between a simulation using the
bitant symmetry (blue dot-dashed line) and one using the full domain
(black solid line).  Clearly, no significant differences can be
observed between the two simulations during the first $25$ ms of
evolution. The same conclusion holds for all quantities related to the
magnetic field and they have been also monitored.

\section{The role of resolution and convergence}
\label{app_b}

\begin{figure}
\includegraphics[width=0.48\textwidth]{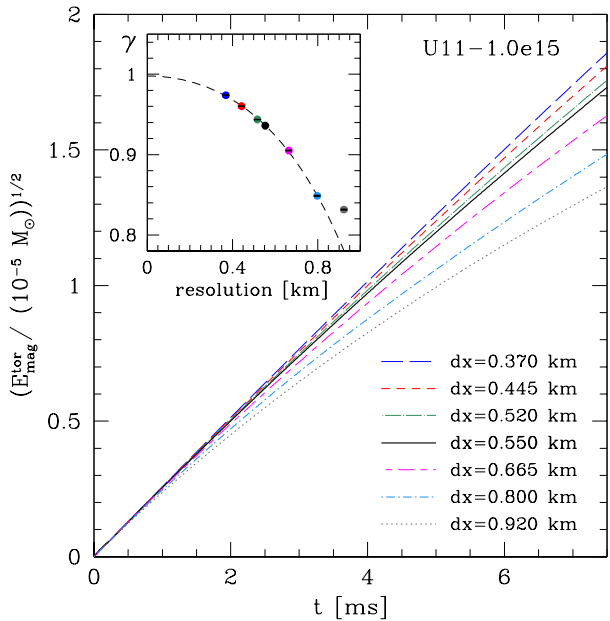}
\caption{Initial growth of the square root of the toroidal component of
  the magnetic energy $E^{\mathrm{tor}}_{\rm{mag}}$
  [Eq. \eqref{eq:EMAG_tor}] for different resolutions of the finest
  grid. The reference resolution, $\Delta x = \simeq 0.550$ km, is shown
  with a black solid line. The inset shows instead the growth rate
  $\gamma$ as a function of the resolution and its fit with a quadratic
  function (dashed line). Note that the expected value $\gamma=1$ is
  approached in the limit of $\Delta x \to 0$.}
\label{fig:grow_Toroidal_res}
\end{figure}

\begin{table}[h]
\begin{tabular}{|cc|c|c|c|c|}
\hline
\hline 
 $\Delta x$  & $\Delta x$ & symmetry & $\eta_{\text{max}}$ & $\tau_{\textrm{bar}}$ & $f_{\textrm{bar}}$  \\
 $[M_\odot]$ &   [km]     & symmetry &  & [ms] &  [Hz] \\
\hline   
  0.250  &  0.370  & bitant  &  $0.743$  &  $1.15^{+0.01}_{-0.01}$ & $491^{+1}_{-1}$ \\
  0.350  &  0.445  & bitant  &  $0.746$  &  $1.16^{+0.03}_{-0.03}$ & $492^{+1}_{-1}$ \\
  0.375  &  0.520  & bitant  &  $0.751$  &  $1.17^{+0.04}_{-0.05}$ & $491^{+2}_{-4}$ \\
  0.375  &  0.520  & full    &  $0.753$  &  $1.14^{+0.01}_{-0.01}$ & $490^{+3}_{-2}$ \\
  0.450  &  0.665  & bitant  &  $0.745$  &  $1.18^{+0.03}_{-0.05}$ & $489^{+2}_{-2}$ \\
  0.540  &  0.800  & bitant  &  $0.754$  &  $1.19^{+0.05}_{-0.05}$ & $487^{+3}_{-5}$ \\
  0.625  &  0.920  & bitant  &  $0.743$  &  $1.20^{+0.11}_{-0.05}$ & $484^{+2}_{-7}$ \\
\hline
\hline
\end{tabular}
\caption{Main properties of the bar-mode instability for model
  \texttt{U11-1.0e15} at different resolutions. Here we report the
  resolution in terms of solar masses and kilometers, the symmetry we
  imposed to the computational domain, the maximum value of the
  distortion parameter $\eta$, the growth times $\tau_{\textrm{bar}}$ and
  the frequencies $f_{\textrm{bar}}$ of the bar-mode deformation.}
\label{table:resolution}
\end{table}

Determining the convergence properties of our simulations is of course
an essential validation of the results presented and a considerable
effort has been put into performing these measures within the
numerical setup used here. Lacking an analytic solution that describes
the fully nonlinear development of the bar, we can only perform
self-convergence tests at this stage. The results will be discussed
below.

However, there is a regime in our calculations in which we can exploit
the knowledge of an analytic solution and this refers to the initial
shearing of the poloidal magnetic field by the differentially rotating
star. It is in fact not difficult to show that within an ideal-MHD
framework the induction equation predicts a growth of the toroidal
magnetic field which is linear in time (see, for instance,
\cite{Rezzolla00} for a pedagogic presentation of the perturbed induction
equation). To explore this regime we have performed a large number of
simulations of model \texttt{U11-1.0e15} with varying resolution and
monitored the growth of the square root of the toroidal magnetic energy
$E^{\mathrm{tor}}_{\rm{mag}}$ [\cf Eq.~(\ref{eq:EMAG_tor}]; we recall
that the poloidal magnetic field is not expected to grow during this
stage (\cf Sect.~\ref{subsec:UntableModels}).

Figure~\ref{fig:grow_Toroidal_res} reports the results of these
simulations relatively to the first $\sim 7$ ms, with different curves
referring to different resolutions. It is then evident that the curves are
getting closer and closer to straight lines as the resolution
increases. To measure whether a linear-in-time-growth is actually
reached we have actually computed the growth rate ``$\gamma$'' by
fitting the square root of the magnetic energy with a trial function
which is a power-law in time with undetermined growth rate, \ie with
\begin{equation}
\sqrt{E^{\mathrm{tor}}_{\rm{mag}}(t)} = y(t)=y_0 + m \,t^\gamma \,, 
\end{equation}
where the time interval has been selected to be between $0.2$ to $5$
ms. 

\begin{figure}
 \includegraphics[width=0.48\textwidth]{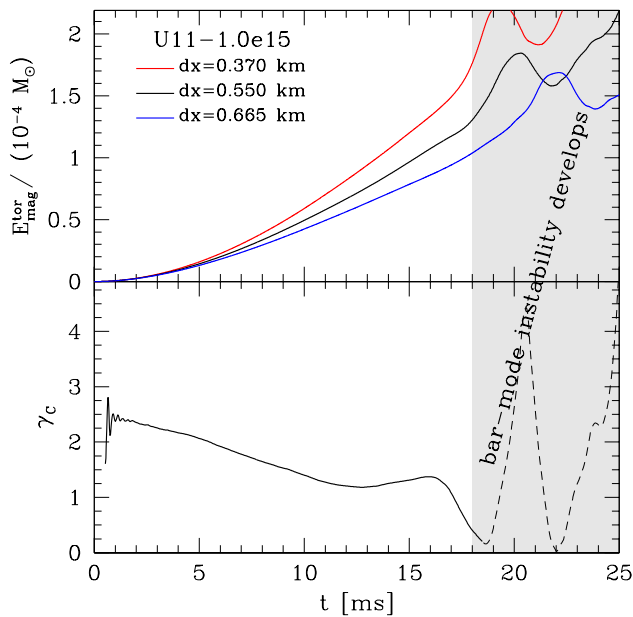}
\caption{\textit{Top panel:} evolution of the toroidal component of the
  magnetic energy $E^{\mathrm{tor}}_{\rm{mag}}$ for three different
  resolutions. \textit{Top panel:} order of the self-convergence test,
  $\gamma_c$, shown as a function of time. Note that a convergence order
  around $2$ is measured before the bar-mode instability develops and
  shocks are produced (gray-shaded area).}
\label{fig:grow_Toroidal_res_2}
\end{figure}

Also reported in the inset of Fig.~\ref{fig:grow_Toroidal_res} are the
values of $\gamma$ (colored symbols) as a function of the resolution
$\Delta x$, as well as a fit for $\gamma(\Delta x)$ (dashed line) when
assuming a second-order convergence with resolution, \ie assuming
$\gamma(\Delta x)=\gamma|_{\Delta x = 0}+k \, \Delta x^2$ (the point
for $\Delta x=0.920$ km has been excluded from the fit). Having made
this assumption, we do find that the growth rate is in very good
agreement with the one expected in this linear regime, with
$\gamma|_{\Delta x = 0} = 1 \pm 0.005$. Of course this result does not
prove directly that we have second-order convergence over this period
of time. However, what it does prove is that if a second-order
convergence is assumed, then our solution matches the expected
perturbative one.

Next we consider a more general calculation of the convergence order
by performing again simulations of model \texttt{U11-1.0e15} for a
range of resolutions. This time our results for the convergence are
obtained by taking into account the data corresponding to the
whole timescale of the simulations, \ie $\sim 25$ ms.  Also
in this case we monitor the growth of the toroidal magnetic energy
$E^{\mathrm{tor}}_{\rm{mag}}$ and report in the top panel of
Fig.~\ref{fig:grow_Toroidal_res_2} its evolution for three runs at
resolutions: $\Delta x = 0.370,\,0.550$, and $0.665$ km, respectively.
The bottom panel of the same figure reports instead the convergence
order $\gamma_c$, computed via a self-convergence
test~\cite{Rezzolla_book:2013}, when shown as a function of time.

In this case it is then possible to recognize that the code does indeed
converge at around second order during the linear growth stage (\ie for
$t \lesssim 5$ ms), in agreement with the results found in purely
hydrodynamical simulations~\cite{Baiotti:2009gk}, or with the new
resistive code~\cite{Dionysopoulou:2012pp}. However, as the bar-mode
instability develops, the second-order convergence is lost and the
convergence order reduces to one. This is not surprising as the
development of the bar also leads to the formation of shocks, which
necessarily degrade our solution to a first-order convergence. We also
note that the large variations in the convergence order shown in the
gray-shaded area of Fig.~\ref{fig:grow_Toroidal_res_2} (\ie for $t \gtrsim
18$ ms) are simply the consequence of the fact that the instability
starts growing at different times for different resolutions and this
inevitably leads to large excursions in $\gamma_c$. Because all the
major considerations made about the onset and development of the bar
deformation, as well as the estimates for the growth rates and
frequencies, are obtained after looking at the first 20 ms of the
evolution, we conclude that all of our results have been achieved with
solutions converging at the expected rates.



\end{document}